\documentstyle[epsfig]{ar}

\begin{document}

\def\beeq{\begin{eqnarray}} \def\eeeq{\end{eqnarray}}
\newcommand\mysection{\setcounter{equation}{0}\section}
\renewcommand{\theequation}{\thesection.\arabic{equation}}
\newcounter{hran} \renewcommand{\thehran}{\thesection.\arabic{hran}}

\def\bmini{\setcounter{hran}{\value{equation}}\refstepcounter{hran}\setcounter{equation}{0}\renewcommand{\theequation}{\thehran\alph{equation}}\begin{eqnarray}}
\def\bminiG#1{\setcounter{hran}{\value{equation}}\refstepcounter{hran}\setcounter{equation}{-1}\renewcommand{\theequation}{\thehran\alph{equation}}\refstepcounter{equation}\label{#1}\begin{eqnarray}}

%
%

\def\emini{\end{eqnarray}\relax\setcounter{equation}{\value{hran}}\renewcommand{\theequation}{\thesection.\arabic{equation}}}

\def\ben{\begin{enumerate}}  \def\een{\end{enumerate}}
\def\bit{\begin{itemize}}    \def\eit{\end{itemize}}
\def\beq{\begin{equation}}   \def\eeq{\end{equation}}
\def\bea{\begin{eqnarray}}  \def\eea{\end{eqnarray}}
\def\nn{\nonumber}
\def\noi{\noindent}


\title{Energy loss in perturbative QCD {\footnote{preprint BI-TP 2000/08,
 LPT-Orsay 00-22 \\ submitted to 
Annual Review of Nuclear and Particle Science}} 
}

\author{R.~Baier
\affiliation{
Fakult\"{a}t f\"{u}r Physik, Universit\"{a}t Bielefeld \\
D-33501 Bielefeld, Germany}
 D.~Schiff
\affiliation{LPT, Universit\'e Paris-Sud, B\^atiment 210 \\
 F-91405 Orsay, France}
B.~G.~Zakharov
\affiliation{L.D.Landau Institute for Theoretical Physics \\
 117334 Moscow, Russia}}

\begin{keywords}
QCD, dense nuclear matter, quark-gluon plasma,
 gluon radiation, energy loss, relativistic heavy ion collisions
 
\end{keywords}

\begin{abstract}
We review the properties of energetic parton propagation in hot or cold QCD
matter, as obtained in recent works. Advances in understanding 
the energy loss -  collisional and  radiative - are summarized,
with emphasis on the latter: 
it features very interesting properties which may help to detect
 the quark-gluon plasma produced in heavy ion collisions.
We describe two different theoretical approaches,
which lead to the same radiated gluon energy spectrum.
The case of a longitudinally expanding QCD plasma is investigated.
The energy lost by a jet with given opening angle is calculated in view
of making predictions for the suppression (quenching) of hard jet production.
Phenomenological implications for the difference between hot and cold
matter are discussed.
 Numerical estimates of the loss suggest that it may be si\-gni\-fi\-can\-tly
enhanced in hot compared to cold matter.
\end{abstract}

\maketitle

\section{INTRODUCTION}

Over the past few years, a lot of work has
 been devoted to the propagation
of high energy partons (jets) through hot and cold QCD matter.
 The jet $p_\perp$-broadening
and the gluon radiation induced by multiple scattering, together with the
resulting radiative energy loss of the jet have been
studied.
These studies are extensions to QCD of the
analogous QED problem considered long ago by Landau, Pomeranchuk and Migdal
\cite{Landau,Migdal}.
Recent measurements \cite{Anthony}
(reviewed in \cite{Klein})
confirm the theoretical predictions  in the QED case to good accuracy.

As in QED, coherent suppression of the radiation spectrum takes place
when a parton propagates in a QCD medium.
New and  interesting predictions are found. When a high
energy parton traverses a length $L$ of hot or cold matter, the induced
radiative energy loss is proportional to $L^2$. The energy loss of a high
energy jet in a hot QCD plasma appears to be much larger than in cold nuclear
matter even at moderate temperatures of the plasma, $T \sim$ 200 MeV. 

The order of magnitude of the effect in hot matter compared to the case of cold
nuclear matter may be expected to be large enough to lead to an observable and
remarkable signal of the production of the quark-gluon plasma (QGP). Indeed, it
has been proposed to measure the magnitude
 of ``jet quenching'' in the transverse
momentum spectrum of hard jets produced in heavy ion collisions,
noting that  jet quenching is the manifestation of energy loss as seen
in the suppression and change of shape of the jet spectrum compared 
with hadron data.

This review{\footnote{We concentrate on the
more recent theoretical advances. References to earlier work
are found in the quoted papers.}}
 is organized as follows:
The case of elastic parton scattering giving rise to the collisional energy
loss, especially in hot matter, is presented in section~2.
In section 3, we give the basic elements of the equations and describe the
coherent pattern of the gluon radiative spectrum 
induced by multiple scattering.
We derive the induced energy loss and the jet transverse momentum
broadening in terms of phenomenologically significant quantities. 
Section 4 is devoted to the path integral approach,
which provides another derivation of the induced radiative spectrum.
For heavy ion collisions the case of an expanding QCD plasma 
is more realistic, and therefore in section 5 we consider
 the corresponding energy loss calculation.
In section 6, we investigate  the angular dependence
of the radiative gluon spectrum.
The dependence of the energy loss on the jet cone size is analyzed.
Section 7 is devoted to estimates of the energy loss in hot
 QCD matter and in nuclear matter, and
orders of magnitude are given. A noticeable result
is that the energy loss in the case of a hot QCD medium
is found to be quite collimated.
 Experimental indications are shortly reviewed.
We close with an outlook.

\section{COLLISIONAL ENERGY LOSS IN QCD}

The electromagnetic energy loss of a charged particle in matter is a 
well studied subject \cite{Jackson,Ter}.
Similar mechanisms are responsible for the energy loss of a fast 
quark or gluon (jet) propagating through dense QCD matter. 

In this section we discuss the loss caused by elastic collisions of the 
propagating quark or gluon off the (light) partons forming the dense 
quark-gluon plasma (QGP). In order to understand the characteristic
features we consider in some detail the loss of a  test 
quark $Q$ traversing a plasma with quarks $q$ and gluons $g$ interacting 
elastically as $Qq \rightarrow Qq$ and $Qg \rightarrow Qg$
\cite{Bjorkena}; for a review, see \cite{Thoma}.

The energy loss per unit length depends on the density $\rho_p$ of the 
plasma constituents $p$ (with momentum $k$) and on the 
differential cross section weighted by the energy transfer $\omega = 
E - E^\prime$, where $E (E^\prime)$ is the energy of the incoming 
(scattered) $Q$, 
\begin{equation}  \label{eq:C1}
- \frac{dE}{dz} = \sum_{p=q,g} \, \int \, d^3 k \rho_p (k) \, \int \, d q^2\,
J \omega \frac{d\sigma^{Qp \rightarrow Qp}}{dq^2}  .
\end{equation}
$J$ denotes the flux factor, $q^2$ the invariant (four) momentum transfer. 
Small values of $q^2$ dominate the collisions,
\begin{equation}  \label{eq:C2}
\frac{d\sigma^{Qp \rightarrow Qp}}{dq^2} \simeq C_p \frac{2\pi \alpha_s^2}
{(q^2)^2} , 
\end{equation}
with $C_q = \frac{N_c^2 - 1}{2 N^2_c}, \, C_g = 1$ for $N_c$ colors.
For a QGP in thermal and chemical equilibrium the densities are given by 
\begin{equation}  \label{eq:C3}
\rho_q = \frac{4 N_c N_f}{(2\pi)^3} n_F (k), \, 
\rho_g = \frac{2(N^2_c - 1)}{(2\pi)^3} n_B (k) , 
\end{equation}
in terms of the Fermi-Dirac (Bose-Einstein) distributions $n_F (n_B)$.
Although the factor $\omega$ in (\ref{eq:C1}) improves the Rutherford
singularity of (\ref{eq:C2}), a logarithmic dependence still remains after 
the $q^2$-integration, which has to be screened by medium effects, i.e. 
with a cut-off related to the Debye mass \cite{LeBellac}: 
$-q^2_{{\rm min}} \simeq m^2_D = O (\alpha_s T^2)$. Noting that 
$J \omega \simeq \frac{q^2}{2k}$ (when $E , E^\prime \gg k\simeq O (T)$)
one obtains \cite{Bjorkena,Thoma}
\begin{equation}  \label{eq:C4}
- \frac{dE}{dz} = \pi \alpha^2_s \, \sum_p \, C_p \, \int \, 
\frac{d^3 k}{k} \rho_p (k) \ln \frac{q^2_{{\rm max}}}{q^2_{{\rm min}}}
\simeq \frac{4\pi \alpha^2_s T^2}{3} \left( 1 + \frac{N_f}{6} \right)
\ln \frac{cE}{\alpha_s T} ,
\end{equation}
with $c$ a numerical constant of $O(1)$, and $N_c = 3$. The strong 
coupling constant may be evaluated at the scale $\alpha_s (T)$ for 
high temperature $T$.

Because of the $T^2$ dependence of (\ref{eq:C4}) it has been pointed out by
Bjorken \cite{Bjorkena}
that the collisional loss is proportional to $\sqrt{\epsilon}$, i.e. the 
square root of the QGP energy density, which in leading order in the 
coupling constant is given by
 $\epsilon = 8\pi^2 T^4 (1 + 21 N_f /32) / 15$ \cite{Nieto}.

A proper and consistent treatment of the screening effects of the plasma in 
the low (soft) exchange momentum region of the collisions
is indeed possible in the thermal field-theoretic framework \cite{LeBellac},
using resummed perturbation theory at high temperature. This method has 
been developed by Braaten and Pisarski \cite{Pisarski,Taylor}
and it allows one to calculate the hard thermal loop (HTL) corrections to the 
propagator of the exchanged gluon in the $Qq \rightarrow Qq$,
and the $Qg \rightarrow Qg$ processes. The quantum field-theoretic calculation
of the energy loss of a  quark requires the evaluation of the 
discontinuity of the self-energy diagrams e.g. illustrated in 
Figure \ref{fig:htl2}. 
For the soft momentum exchange (with momentum less than $q_c = O (g^{1/2} T)$)
the HTL gluon propagator contributes, whereas for the hard momentum 
exchange $(q\simeq T)$ the tree-level elastic scattering 
(Figure \ref{fig:htl2}b)
contributes \cite{Braatena}.

\begin{figure}
\begin{center}
(a)
\epsfig{file=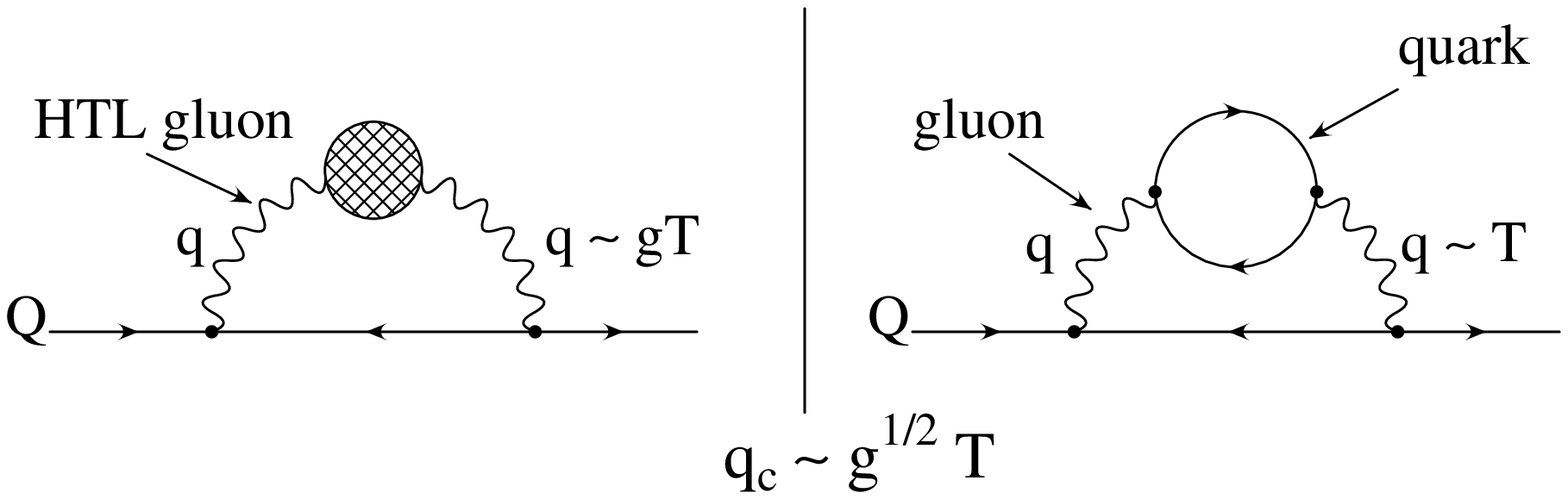, width=12cm}
(b)
\end{center}
\caption{Self-energy  diagrams contributing to the collisional energy loss:
(a) in HTL-resummed 
 perturbation theory for soft exchanged momentum and (b)
  in fixed leading order for hard exchanged momentum.}
\label{fig:htl2}
\end{figure}

The momentum cut-off $q_c$ drops out in the sum of soft and hard contributions.
 It is important to note that 
Landau-damping effects, because of the negative $q^2$ values in (\ref{eq:C1}),
screen successfully the low $q^2$ region leading to a well defined result 
for $dE/dz$ (at least to leading order in the coupling constant).

As an example the result is illustrated in 
Figure \ref{fig:braat},
where the energy loss of a charm quark is shown by the dashed curve using 
parameters characteristic for a thermalized QGP as expected in 
ultrarelativistic heavy ion collisions.

\begin{figure}
\begin{center}
   \epsfig{bbllx=94,bblly=264,bburx=483,bbury=587,
file=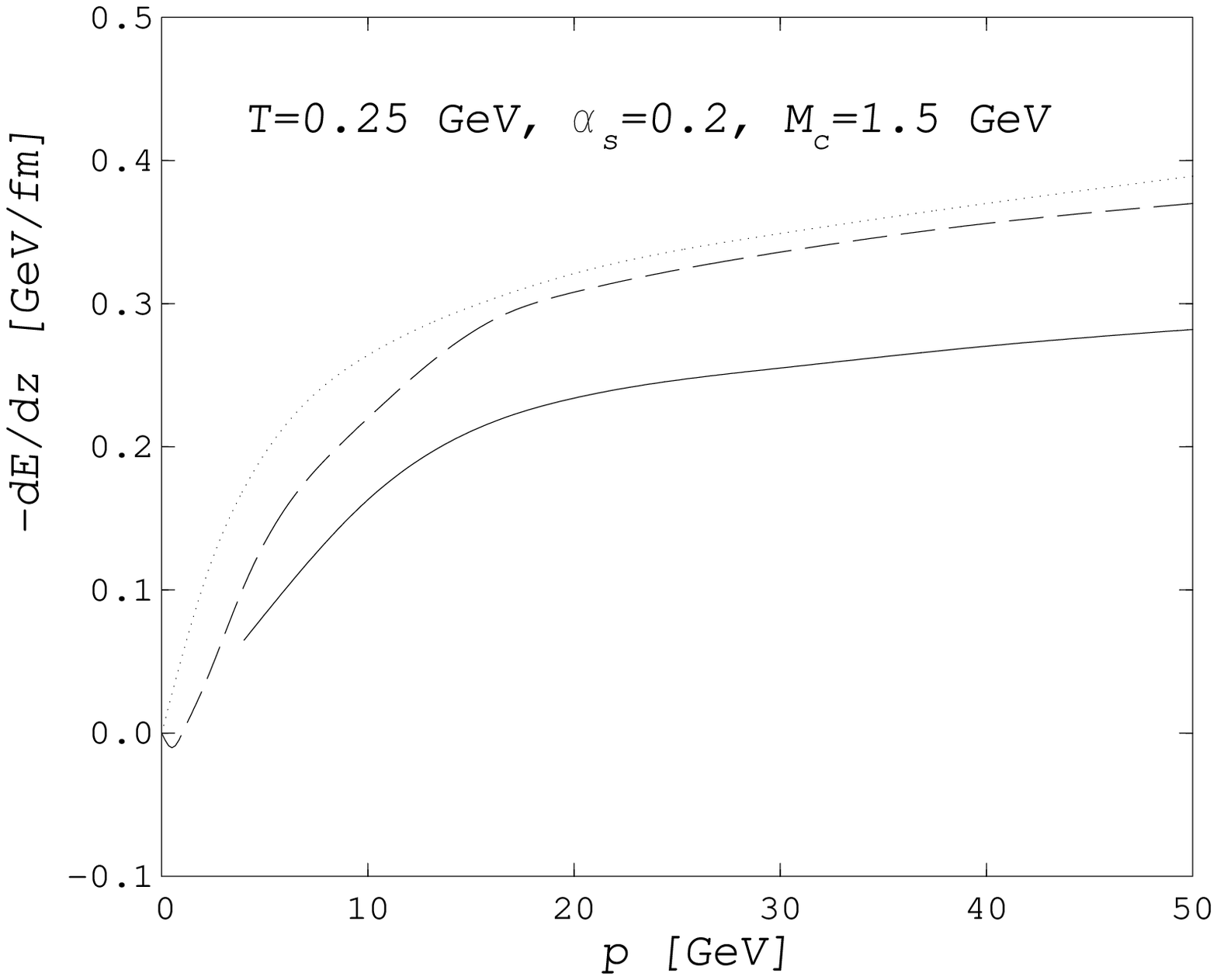, width=70mm,height=60mm}
\end{center}
\vskip 1truecm
\caption{Collisional energy loss
of a charm quark as a function of its momentum. The quark propagates through an
 out-of-chemical equilibrium plasma with fugacities
 $\lambda_g = 1, \lambda_q = 0$ (solid curve) \cite{Dirks}.
The dashed curve is the equilibrium result of \cite{Braatena}, 
the  dotted curve shows the original prediction by Bjorken \cite{Bjorkena}
.}
\label{fig:braat}
\end{figure}

For light quarks the collisional energy loss  for a 
jet propagating in a hot medium of $T = 0.25$ GeV  
amounts to 0.2 - 0.3 GeV/fm  \cite{Thoma,Thomab},
 in agreement with the estimates of  
\cite{Bjorkena} as shown  in Figure \ref{fig:braat} (dotted curve).
For a gluon jet the loss is predicted to be larger by the color factor 
$2N^2_c / (N^2_c - 1) = 9/4$ than for the quark jet. 

Since the QGP expected at RHIC and LHC is likely to be out of
chemical equilibrium it is necessary to investigate the energy loss in 
this case \cite{Mustafa,Dirks}.
Indeed, even away from chemical equilibrium,  dynamical screening 
 remains operational within the HTL-resummed perturbation 
theory. More explicitly, the collisional energy loss for a heavy quark 
(mass $M$) propagating through a QGP parametrized in terms of the 
distribution functions $\lambda_q n_F$  and $\lambda_g n_B$,
respectively,  where 
$\lambda_{q,g}$ are the fugacity factors describing chemical 
non-equilibrium, becomes  
\begin{equation}   \label{eq:C6}
  - \frac{dE}{dz}  = 2 \alpha_s {\tilde m}^2_g 
\ln \left[ 0.920 \frac{\sqrt{ET}}{ \tilde m_g} \,
 2^{\lambda_q N_f/(12 \lambda_g + 2 \lambda_q N_f)} 
                                \right].
\end{equation}
This expression \cite{Dirks} 
is valid for energetic quarks with $E \gg M^2/T$ and 
contains for $\lambda_q = \lambda_g = 1$ the original result of 
\cite{Braatena}.
The screening mass parameter is 
\begin{equation}   \label{eq:C7}
\tilde m^2_g = 4 \pi \alpha_s (\lambda_g + \lambda_q N_f / 6 ) T^2 / 3 .
\end{equation}
For comparison the solid curve in 
Figure \ref{fig:braat}
shows the loss for the interesting case of the ``early plasma phase'' 
which is dominated by gluons ($\lambda_g = 1 , \lambda_q = 0$), where the 
loss is exclusively due to elastic $Qg \rightarrow Qg$ scattering mediated
by gluon exchange. 

In summary, even when the partons propagating in hot matter have a
 large momentum, the 
collisional energy loss per unit length turns out to be less than 
$O (1)$ GeV/fm when reasonable values for $\alpha_s$ and $T$ are taken. 
This estimate may be compared with the value for the hadronic string 
tension, $\kappa \simeq 1$ GeV/fm, which measures the slowing down of a 
high momentum quark in (cold) nuclear matter \cite{Casher}.

\section{RADIATIVE ENERGY LOSS IN QCD}

\subsection{Model, basic parameters and equations}

We imagine a very energetic quark of energy $E$ propagating through a 
QCD medium of finite length $L$. Multiple scattering of this projectile 
in the medium induces gluon radiation, which gives rise to the quark 
energy loss.

 The main assumption \cite{Gyulassy,Wang} is that the
scattering centers are static and uncorrelated (in the spirit of the Glauber
picture). We thus focus on purely
 radiative processes since the collisional energy
loss vanishes in the case of static centers. 

We define a normalized quark-``particle'' cross-section 
\begin{equation}
\label{1e}
V(Q^2) = {1 \over \pi \sigma} \ {d\sigma \over dQ^2} ,
 \end{equation}
 
\noindent where $Q$ is the 2-dimensional transverse momentum transfer scaled
by an appropriate scale~:
\begin{eqnarray}
\label{2e}
& &\vec{Q} = {\vec{q} \over \mu} , \nonumber \\
&\hskip - 4.5 truecm \hbox{and} \hskip 4 truecm &\sigma = \int {d \sigma \over
d^2Q} \ d\vec{Q} \quad . \end{eqnarray}

\noindent In the case of a hot QCD plasma, the ``particle'' is a quark or
gluon and it is a nucleon in the case of cold matter. 
$d\sigma/d^2 Q$ depends only on $\vec{Q}$, 
as it is usually assumed for diffractive kinematics with very large incident
energy. 
The scale $\mu$ characteristic of the medium is conveniently taken
as the Debye screening mass in the hot case and as a typical momentum
transfer in a quark-nucleon collision. The condition that the independent
scattering picture be valid may be expressed as~:     
\begin{equation}
\label{3e}
\mu^{-1} \ll \lambda ,
\end{equation}
\noindent where $\lambda$ is the parton mean free path in the medium $\lambda =
1/\rho \sigma$, and $\rho$ is the density of the medium.
 We assume that a large number of scatterings takes place, that is
\begin{equation}
\label{4e}
L \gg \lambda . 
\end{equation}

Successive  scatterings being independent, the parton propagation is 
"time-ordered" and time-ordered perturbation theory is the natural
framework to calculate the radiation amplitude. Let us give a  sketch 
of the basic equations, referring the reader to \cite{Baier1,Baier2}
 for further details. 
We may number the scattering centers depending on the interaction time
and write for the radiation spectrum induced by $N$ scatterings
\beq\label{SpectrumiN}
 \omega {dI \over d \omega} = {\alpha_s \over 2\pi^2} \int 
d \vec{k}_{\bot} \left < \sum_{i=1}^N \sum_{j=1}^N {\vec{J}^i_{eff} \cdot 
\vec{J}^{j\dagger}_{eff}}
 \ e^{i(\varphi_i - \varphi_j)}  \right > \> , 
\eeq 
where $\vec{J}^i_{eff}$ is an effective current for the gluon  emission
induced by center $i$. It includes color factors consistent with the
overall normalization to the elastic scattering cross section.
The phase $\varphi_i$ is associated to time $t_i$ (longitudinal coordinate
$z_i$) by $\varphi_i = t_i k_{\bot}^2/\omega$.

The brackets indicate averaging -  over 
momentum transfers and over $z_i$  - for which a simplified model is 
\beq
\left < (\ \ ... \ \ ) \right >
 \Leftrightarrow \int \prod_{\ell = 1}^{N-1} {d \Delta_{\ell}
\over \lambda} \exp \left ( - {\Delta_{\ell} \over \lambda} \right ) \cdot \int
\prod_{i=1}^N d \vec{q}_{i_{\bot}} \ V(q_{i_{\bot}}^2) (\ \ ... \ \  ) \
,\label{[Average]}  
\eeq
where $\Delta_{\ell} = z_{\ell + 1} - z_{\ell}$.
We  rewrite (\ref{SpectrumiN}) as
\begin{eqnarray} \label{Spectrum2}
  \omega {dI \over d\omega} &=& 
{\alpha_s \over 2 \pi^2} \int d \vec{k}_{\bot}
\nonumber \\
  &\cdot &  \left < 2 \ {\rm Re} \sum_{i=1}^N \sum_{j=i+1}^N  
 \vec{J}^i_{eff} \cdot \vec{J}^{j\dagger}_{eff} 
  \left ( e^{i (\varphi_{_i} - \varphi_{_j})} - 1\right ) 
 + \left | \sum_{i=1}^N \vec{J}^i_{eff} \right |^2 \right > ,   
\end{eqnarray}
which
allows one to exhibit the so-called factorization contribution: 
the second term in (\ref{Spectrum2})  which
corresponds to the limit of vanishing phases. 
This contribution is  equivalent to the radiation spectrum induced 
by a single scattering of momentum transfer 
$\vec{q}_{\bot_{tot}} = \sum\limits_{i=1}^N \vec{q}_i$.
It has at most a weak logarithmic medium dependence \cite{Baier2}.
We concentrate in the following 
 on the medium-induced radiation spectrum and drop the factorization  term.
In the limit of large-$N_c$ 
and replacing sums over $i$ and $j$ by integrals, i.e. 
taking the sum over scatterings to be arbitrary in number,
the following expression for the spectrum
 per unit length of the medium
is obtained,

\[
\omega {dI \over d \omega dz} = { \alpha_s N_c \over 2 \pi^2 L}
 \int_0^L d \Delta
\int_0^{L- \Delta} {dz_1 \over \lambda} \int d \vec{U}  
\]
\beq\label{spec2}
\cdot \left < 2 {\rm Re}
\sum_{n=0}^{\infty} \vec{J}_1 \cdot \vec{J}_{n+2} \left[\exp\left\{ i\kappa 
\sum_{\ell = 1}^{n+1} U_{\ell}^2 {\Delta_{\ell} \over \lambda}
 \right \} - 1 \right ] 
\delta  \left (\Delta - \sum_{m=1}^{n+1} \Delta_m \right ) \right > , 
\eeq

\noi
with $\vec{U} = \vec{k}_{\bot}/\mu$ and $\kappa = \lambda \mu^2/2 \omega$.
In the soft gluon limit the  rescaled emission current is given by
\beq
\vec{J}_i  = \left ( {\vec{U}_i \over U_i^2} - {\vec{U}_i + \vec{Q}_i
\over ( \vec{U}_i + \vec{Q}_i )^2} \right ) . 
\label{4.4}
\eeq 

\noi
Eq. (\ref{spec2}) exhibits the interference nature and coherent pattern
of the spectrum. The phase factor as it appears here may be understood 
in terms of formation time arguments which will be discussed heuristically
in the next section. It can be shown moreover \cite{Baier2}
that (\ref{spec2}) leads to a simple structure of the spectrum
per unit length which will be the starting point of section 3.4:
\beq \label{4.7}
\left . \omega {dI \over d \omega dz} = {\alpha_s N_c \over {\pi^2 L}}
 \, {\rm Re} \int_0^L {d \Delta \over \lambda}
\int_0^{L- \Delta} {dz_1 \over \lambda} 
\int {d \vec{U}} \vec{f}(\vec{U}, \Delta)
\cdot \vec{f}_{Born}(\vec{U}) 
 \right |_{\kappa}^{\kappa = 0} \> ,
\eeq 
where $\vec{f}_{Born}(\vec{U})$ is the {\underline{Born amplitude}}
defined  as
\beq\label{4.6} 
\vec{f}_{Born}(\vec{U}) =  \int d \vec{Q}_{1} \ V(Q_{1}^2) \ 
\vec{J}_{1} \>,
\eeq
and $\vec{f}(\vec{U}, \Delta)$ is the {\underline{evolved amplitude}}
which satisfies a Bethe-Salpeter type equation. Subtracting in 
(\ref{4.7}) the contribution for $\kappa =0$ corresponds in (\ref{spec2})
to subtracting the zero phase contribution.
The generic structure of (\ref{4.7})
is illustrated in Figure \ref{fig:cohere}.
The reader may question the one-gluon exchange approximation as shown in
this Figure \ref{fig:cohere}. In fact, the scale of the coupling for
each individual scattering is set by the accumulated overall transverse
momentum. This justifies the perturbative treatment.

\begin{figure}[ht]
\centering
\epsfig{file=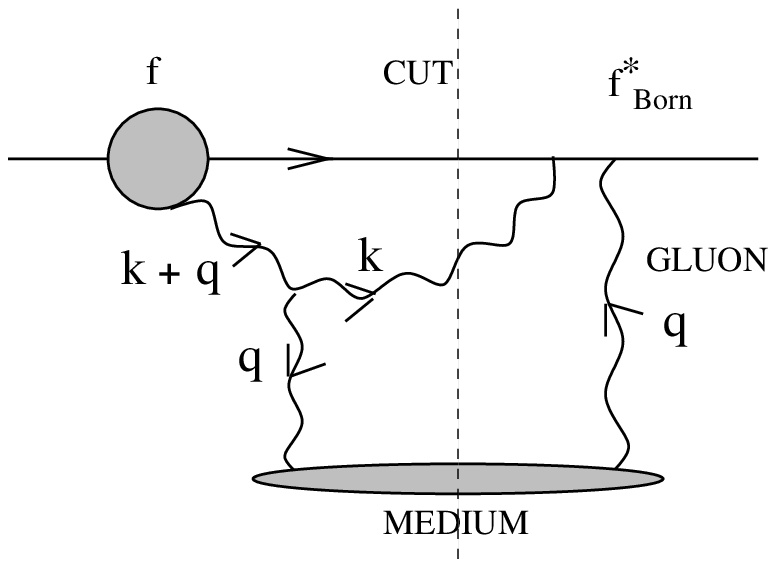,angle=-90, width=7cm}
\vskip 2cm
\caption{\label{fig:cohere}
Contribution to the induced gluon spectrum by interference
between the amplitude $f$ and the Born amplitude $f_{Born}$
. }
\vskip 1.5cm
\end{figure}

\subsection{Heuristic discussion}

In the following discussion we neglect logarithmic and numerical factors of 
$O(1)$, but keep all relevant parameters.

The semi-quantitative argument which allows one to understand the coherent 
pattern of the induced gluon radiation is the following. One defines the 
formation time of the radiation, 
\begin{equation} \label{eq:R2}
t_{{\rm form}} \simeq \frac{\omega}{k_\bot^2},
\end{equation}
where $\omega$ and $k_\bot$ are the gluon energy and transverse momentum 
(with $\omega \gg k_\bot$ and the typical $k_\bot \simeq \mu$). 
When $t_{{\rm form}} \gg \lambda$, radiation takes place in a coherent 
fashion with many scattering centers acting as a single one.
Let us introduce the  coherence time (length) $l_{{\rm coh}}$
which plays an  important role in the following considerations. 
It is associated with the formation time of a gluon radiated by
a group of scattering centers which acts as 
\underbar{one} source of radiation,
\begin{equation}  \label{eq:R3}
l_{{\rm coh}} \simeq \frac{\omega}{\langle k_\bot^2 \rangle_{l_{{\rm coh}}}}, 
\end{equation}
with
\begin{equation}  \label{eq:R4}
\langle k_\bot^2 \rangle_{l_{{\rm coh}}} \simeq \frac{l_{{\rm coh}}}{\lambda} 
\mu^2 \equiv N_{{\rm coh}} \mu^2 , 
\end{equation} 
assuming a random walk expression for the accumulated  gluon 
transverse momentum. 
One derives the estimate
\begin{equation}  \label{eq:R5}
l_{{\rm coh}} \simeq \sqrt{\frac{\lambda}{\mu^2} \omega},
\end{equation}
so that 
the number of coherent scatterings becomes
\begin{equation}  \label{eq:R5N}
 N_{{\rm coh}} \simeq \sqrt{\frac{
\omega}{\lambda\mu^2}} \equiv \sqrt{\frac{\omega}{E_{{\rm LPM}}}} ,
\end{equation}
where the energy parameter $E_{{\rm LPM}} \equiv \lambda \mu^2$ is 
introduced \cite{Dokshitzer}, in analogy with the QED
Landau-Pomeranchuk-Migdal (LPM) phenomenon.

For small $\omega \le E_{{\rm LPM}}$, incoherent radiation takes
place on $L/\lambda$ scattering centers.
Using the soft $\omega$ limit for the single scattering spectrum 
\cite{Gunion} 
\begin{equation}  \label{eq:R6}
\frac{\omega dI}{d\omega} \simeq \frac{\alpha_s}{\pi} N_c , 
\end{equation}
 the differential energy spectrum per unit length in the so-called
Bethe-Heitler (BH) regime for incoherent radiation is derived,
\begin{equation}  \label{eq:R7}
\frac{\omega dI}{d\omega\, dz} {\Bigg|}_{{\rm BH}} = \frac{1}{L} \,
\frac{\omega dI}{d\omega} {\Bigg|}_L \simeq \frac{\alpha_s}{\pi} N_c 
\frac{1}{\lambda} , 
\end{equation} 
with $l_{{\rm coh}} \le \lambda$ and
$\omega \le \omega_{{\rm BH}} \equiv E_{{\rm LPM}}$.

The interesting regime of coherent radiation (\underbar{LPM regime}) is 
defined by 
 $\lambda < l_{{\rm coh}} < L$ ($N_{{\rm coh}} > 1$), i.e.
\begin{equation}  \label{eq:R8}
E_{{\rm LPM}} = \omega_{{\rm BH}} < \omega < \, {\rm min} \, \{ 
\omega_{{\rm fact}} , E \} , 
\end{equation}
with  $\omega_{{\rm fact}} \sim \frac{\mu^2}{\lambda} L^2$.
Since the $N_{{\rm coh}}$ groups are acting as effective single 
scattering centers the energy spectrum is estimated as 
\begin{equation}    \label{eq:R9}
\frac{\omega dI}{d\omega \, dz} {\Bigg|}_{{\rm LPM}} \simeq \frac{1}
{l_{{\rm coh}}} \, 
\frac{\omega dI}{d\omega} {\Bigg|}_{l_{{\rm coh}}} \simeq 
\frac{\alpha_s}{\pi} N_c \frac{1}{l_{{\rm coh}}} \simeq \frac{\alpha_s}{\pi}
N_c \sqrt{\frac{\mu^2}{\lambda} \frac{1}{\omega}}. 
\end{equation}  
Comparing (\ref{eq:R9}) with (\ref{eq:R7}) we find   
a suppression factor given by  $\sqrt{E_{{\rm LPM}} / \omega}$. 

For $l_{{\rm coh}} \ge L$, i.e. when
\begin{equation}   \label{eq:R10}
\omega > \omega_{{\rm fact}}
 = E_{{\rm LPM}} \left( \frac{L}{\lambda} \right)^2 , 
\end{equation}
effectively only one scattering is active (\underbar{factorization 
regime}), and correspondingly for $\omega_{{\rm fact}} < \omega < E$, 
\begin{equation}   \label{eq:R11} 
\frac{\omega dI}{d\omega \, dz} {\Bigg|}_{{\rm fact}} \simeq 
\frac{\alpha_s}{\pi} N_c \frac{1}{L}. 
\end{equation} 

The expressions for the spectrum per unit length
 in the different regimes (Eqs.(\ref{eq:R7}),
(\ref{eq:R9}) and (\ref{eq:R11})) hold for a medium of finite length
$L < L_{cr}$, where 
\begin{equation}  \label{eq:R12}
L_{cr} = \lambda \sqrt{E / E_{{\rm LPM}}}, 
\end{equation}
as derived from the condition that  $\omega_{{\rm fact}} \le  E$ 
(correspondingly, the condition $E > E_{cr} = E_{{\rm LPM}} (L / \lambda)^2$
has to be satisfied). A discussion of the radiation spectrum can be 
found in \cite{Baier2}. 

In order to obtain the energy loss per unit distance $ - dE / dz$ one 
integrates the gluon spectrum over $\omega$, with $0 \le \omega \le E$. 
In addition to a medium independent contribution proportional to 
$\frac{\alpha_s}{\pi} N_c \frac{E}{L}$ (the factorization contribution), we 
obtain from (\ref{eq:R9}) the  medium induced (LPM) loss,  proportional to 
the size of the medium and given by 
\begin{equation}  \label{eq:R13}
- dE / dz \simeq \frac{\alpha_s}{\pi} N_c \sqrt{\frac{\mu^2}{\lambda}
\omega_{{\rm fact}}} \simeq \frac{\alpha_s}{\pi} N_c \frac{\mu^2}{\lambda} 
L , 
\end{equation}
for $L < L_{cr}$. Integrating over $z$ leads to the total loss growing as 
$L^2$.
  For 
$L > L_{cr}$ (i.e. $E < E_{cr}$), the size does not affect the loss per 
unit length, 
\begin{equation}   \label{eq:R14}
- dE / dz \simeq \frac{\alpha_s}{\pi} N_c \sqrt{\frac{\mu^2}{\lambda} E} = 
\frac{\alpha_s}{\pi} \, \frac{N_c}{\lambda} \sqrt{E_{{\rm LPM}} E} , 
\end{equation} 
i.e. a dependence proportional to $\sqrt E$ is obtained, which is familiar 
from the QED-coherent LPM suppression \cite{Landau}.

\begin{figure}[ht]
\centering
\epsfig{file=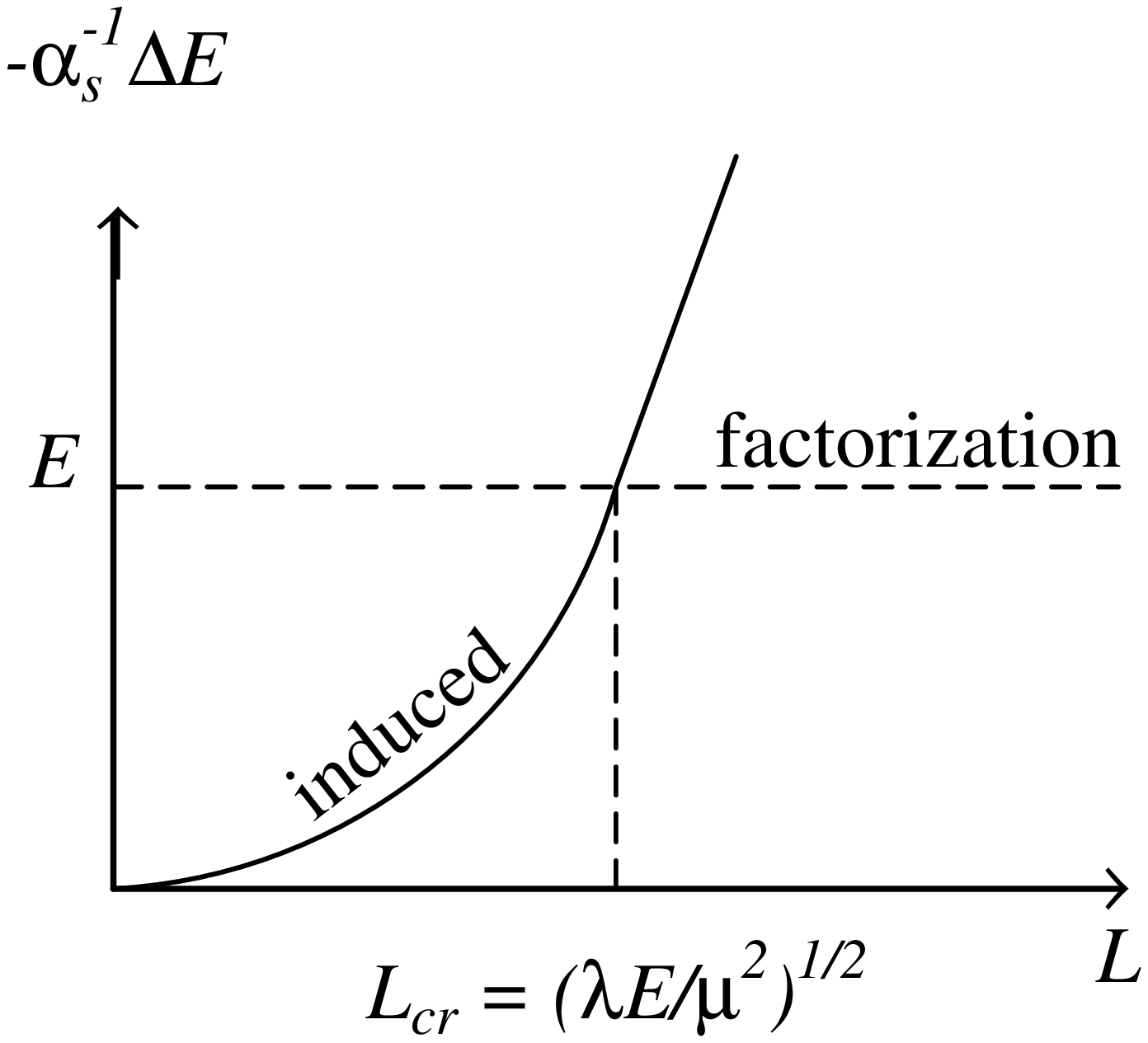, width=7cm, height=5cm}
\caption{Energy loss as a function of the medium size $L$.}
\label{fig:figloss}
\end{figure}

In Figure \ref{fig:figloss}
 the  energy loss 
\begin{equation}   \label{eq:Rloss}
 -\Delta E \equiv \int_0^L   - \frac{dE}{dz} dz, 
\end{equation}
for the induced and the 
factorization cases, is shown as a function of $L$
($N_c$ is taken to be 1).

Using the random walk expression
for the accumulated transverse momentum
of the gluon  due to successive scatterings 
in the medium of size $L$,
\begin{equation}   \label{eq:R15}
\langle k_\bot^2 \rangle_L \simeq \mu^2 L/\lambda , 
\end{equation}
and inserting  this relation in (\ref{eq:R13}), one  obtains 
\begin{equation}  \label{eq:R16}
- dE / dz \simeq \frac{\alpha_s}{\pi} N_c \langle k_\bot^2 \rangle_L , 
\end{equation} 
a relation between the 
induced energy loss and the jet broadening,
which is independent of the details of the interaction.

\subsection{Jet $p_\perp$-broadening}

On the way to deriving the gluon radiative spectrum, let us start with the
classical diffusion equation satisfied by the transverse momentum distribution
of a high energy parton which encounters multiple scattering in a medium.
Suppose the parton is produced in a hard collision with an initial transverse
momentum distribution $f_0(U^2)$~; $U$ is the dimensionless transverse momentum
$\vec{U} \equiv {\vec{p}  / \mu}$ and $\int d \vec{U} f_0(U^2) = 1$. \par

Neglecting the transverse momentum given to the parton by 
induced gluon emission,
one can derive a  kinetic equation for the transverse momentum
distribution $f(U^2, z)$ after a distance $z$ in the medium 
\cite{Baier3}. \par

In terms of the variable $t = {z / \lambda_R}$ with $\lambda_R$ the mean
free path for a parton of color representation $R$, one finds the following
gain-loss equation
\begin{eqnarray}
\label{7e}
{\partial f(U^2, t) \over \partial t} =  &+&  \int f(U'^2, t) \ V((\vec{U}' -
\vec{U})^2) d \vec{U}^{\prime} \nonumber \\
&-& \int f(U^2, t) \ V((\vec{U} - \vec{U}')^2) d \vec{U}^{\prime},
\end{eqnarray}
with
 \begin{equation}
\label{8e}
    f(U^2, 0) = f_0 (U^2) .
\end{equation}

\noi
Defining  $\widetilde{f}(B^2, t)$ as
\begin{equation}
\label{9e}
\widetilde{f}(B^2,t) = \int d \vec{U}
 \ e^{-i \vec{B}\cdot \vec{U}} \ f(U^2, t),
\end{equation}
\noindent and
\begin{equation}
\widetilde{V}(B^2) = \int d \vec{Q} \ e^{-i\vec{B}\cdot \vec{Q}} \ V(Q^2) ,
\label{10e}
\end{equation}

\noindent we find
\begin{equation}
\label{11e}
{\partial \widetilde{f}(B^2, t) \over \partial t} = - {1 \over 4} B^2
\widetilde{v}(B^2) \widetilde{f}(B^2, t) , \end{equation}

\noindent with
\begin{equation}
\label{12e}
\widetilde{v}(B^2) \equiv {4 \over B^2} (1 - \widetilde{V} (B^2)) .
\end{equation}

\noindent
It is possible to define
 a characteristic width of the
distributions $f(U^2, t)$ which is found to be \cite{Baier3}:
\begin{equation}
\label{21e}
p_{\bot W}^2 = {\mu^2 \over \lambda_R} L \ \widetilde{v}(\lambda_R /L)  .
\end{equation}

\noindent The linear growth with $L$ is expected and is used to discuss
$p_{\bot}$-broadening of high energy partons in nuclei. The coefficient
$\widehat{q} = {\mu^2 \over \lambda} \widetilde{v}$ plays the role
of a transport coefficient as encountered in diffusion equations.
(\ref{21e}) is valid for hot and cold
QCD media.

\subsection{The radiative gluon energy spectrum and induced energy loss}

Let us now turn to the gluon spectrum
{\footnote{For related discussions of gluon bremsstrahlung in dense matter
see also \cite{Knoll,Kovchegov}}}. We shall concentrate on a
quark jet. The general case is given in \cite{Baier4}.
 As sketched in section 3.1 and derived in \cite{Baier1,Baier2,Baier4},
 the spectrum
for the radiated gluon is calculated in terms of the interference term between
the quark-gluon amplitude at time $t$ and the complex conjugate Born amplitude.
For simplicity, we restrict ourselves 
here to the case where the quark enters the medium
from outside. (An additional term is needed  in the case when the
quark is produced via a hard scattering at $t = 0$ in the medium). We denote by
$f(\vec{U}, \vec{V}, t)$ the quark gluon amplitude at time $t$. $\vec{U}$ is
the scaled gluon momentum $\vec{U} \equiv {\vec{k} \over \mu}$ and $\vec{V} -
\vec{U}$ the scaled quark momentum as illustrated in Figure \ref{fig:pict5}.
 
\begin{figure}
\begin{center}\mbox{\epsfig{file=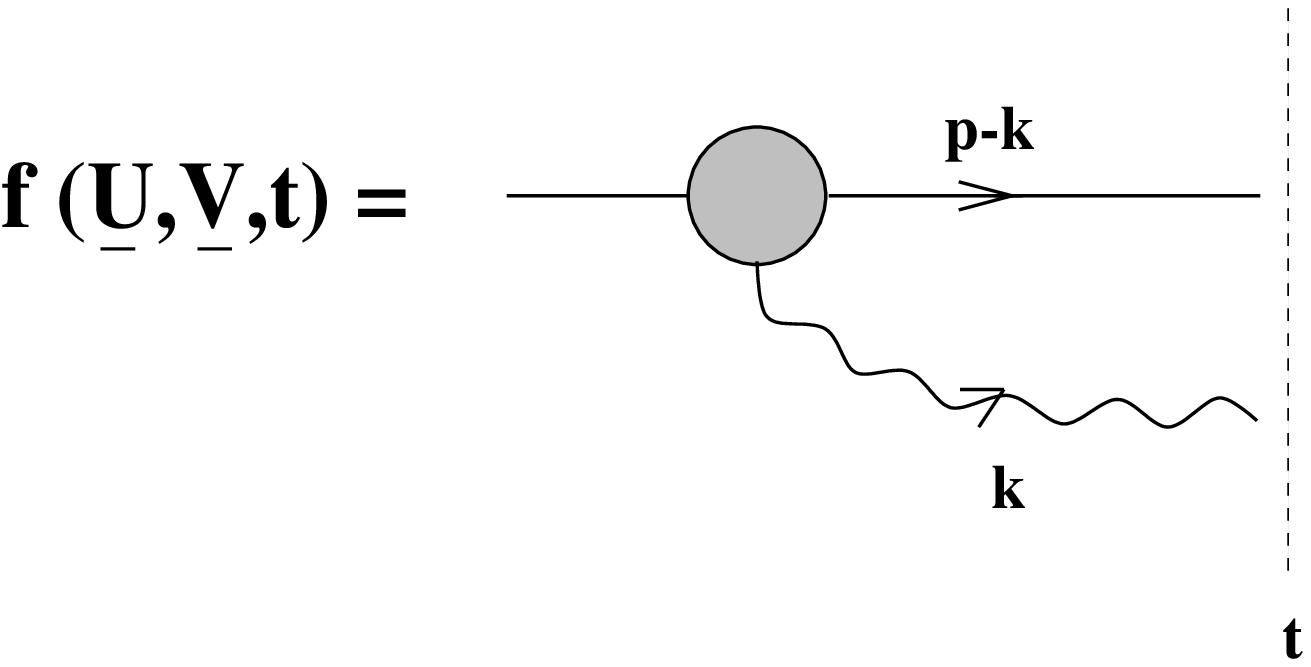,height=4cm,width=8cm}}
\end{center}
\caption{\label{fig:pict5}
Quark gluon amplitude at time $t$.}
\vskip 1truecm
\end{figure}

To account for the gluon
polarization, $f$ is a 2-dimensional vector which will be implied hereafter.
The dependence on $\vec{U}$ and $\vec{V}$ is actually only in the combination
$\vec{U} - x\vec{V}$ with $x = {k / p}$. 
The amplitude  $f(\vec{U}, \vec{V}, t)$
satisfies the initial condition $f(\vec{U}, \vec{V}, 0) = f_{Born}(\vec{U},
\vec{V})$, where $f_{Born}$ is the Born amplitude to be described shortly. \par

The induced gluon spectrum is written as~:
\begin{eqnarray}
&& {\omega \ dI \over d\omega\ dz}  = {\alpha_s \ C_F \over \pi^2 L} 
2 Re \int d \vec{U}
\left \{ \int_0^L dt_2 \int_0^{t_2} dt_1 \right .\nonumber \\ 
 && \cdot \left . \left [ \rho \sigma {N_C \over 2 C_F}
 f (\vec{U} - x \vec{V}, t_2 -
t_1) \right ]  \left [ \rho \sigma {N_C \over 2C_F} f_{Born}^* (\vec{U} - x
\vec{V})\right ] \right \}_{\omega}^{\omega = \infty} .
\label{13e} 
\end{eqnarray}

\noindent The various terms in (\ref{13e}) have simple interpretations. The
${\alpha_S C_F \over \pi^2}$ is the coupling of a gluon to a quark. The $1/L$
comes because we calculate the spectrum per unit length of the medium. The
factor ${N_c \over 2C_F} f(\vec{U} - x \vec{V} , t_2 - t_1) \rho \sigma dt_1$
is the number of scatterers in the medium, $\rho \sigma dt_1$, times the
amplitude with gluon emission at $t_1$, evolved in time up to $t_2$, the time
of emission in the complex conjugate amplitude. The factor ${N_c \over 2C_F}
f_{Born}^*(\vec{U} - x \vec{V})\rho \sigma dt_2$ gives the number of scatterers
times gluon emission in the complex conjugate Born amplitude. The subtraction
of the value of the integrals at
 $\omega = \infty$ e\-li\-mi\-na\-tes the medium independent 
zero-phase contribution. Eq. (\ref{13e}) may be simplified using $t \equiv
\left ( {2 C_F \over N_c} \lambda  \right ) \tau$. Defining $\tau_0 = {N_c
\over 2 C_F} {L \over \lambda}$, we obtain
\begin{eqnarray}
\label{14e}
&& {\omega dI \over d \omega \ dz} =
 {\alpha_s \ N_c \over \pi^2\lambda} Re \int d \vec{Q} 
\nonumber \\   &&\cdot
\left \{ \int_0^{\tau_0} d\tau \left ( 1 - {\tau \over \tau_0} \right )
f(\vec{U} - x \vec{V}, \tau ) \cdot f_{Born}^* (\vec{U} - x \vec{V}) \right
\}_{\omega}^{\omega = \infty} .   
\end{eqnarray}

\noindent Due to the specific dependence of $f$ and $f_{Born}$ on $\vec{U}$ and
$\vec{V}$, it is possible to express them in terms of a single impact parameter
as~:
\begin{eqnarray}
\label{15e}
&&f(\vec{U} - x \vec{V}, \tau ) = \int {d \vec{B} \over (2 \pi )^2}
 \ e^{i \vec{B}
\cdot ( \vec{U} - x \vec{V})} \widetilde{f} (\vec{B}, \tau ) , \nonumber \\
&&f_{Born}(\vec{U} - x\vec{V})
 = \int {d \vec{B} \over (2 \pi )^2} \ e^{i \vec{B} \cdot
( \vec{U} - x \vec{V})} \widetilde{f}_{Born} (\vec{B}) ,
 \end{eqnarray}
\noindent allowing us to obtain the following expression for the spectrum in
impact parameter space~:
\begin{equation}
\label{16e}
{\omega dI \over d \omega \ dz} = {\alpha_s N_c \over 2 \pi^3 \lambda} Re \int
{d \vec{B} \over 2 \pi}  \left \{ \int_0^{\tau_0}
 d\tau \left ( 1 - {\tau \over \tau_0}
\right )  \widetilde{f} (\vec{B}, \tau)
 \cdot \widetilde{f}_{Born}^* (\vec{B})
\right \}_{\omega}^{\omega = \infty} \ .
 \end{equation}

\noi
The generic diagram appears in Figure \ref{fig:cohere}. The complete list
of diagrams describing the Born amplitude is shown in Figure \ref{fig:pict2}.
\begin{figure}[ht]
\centering
\epsfig{file=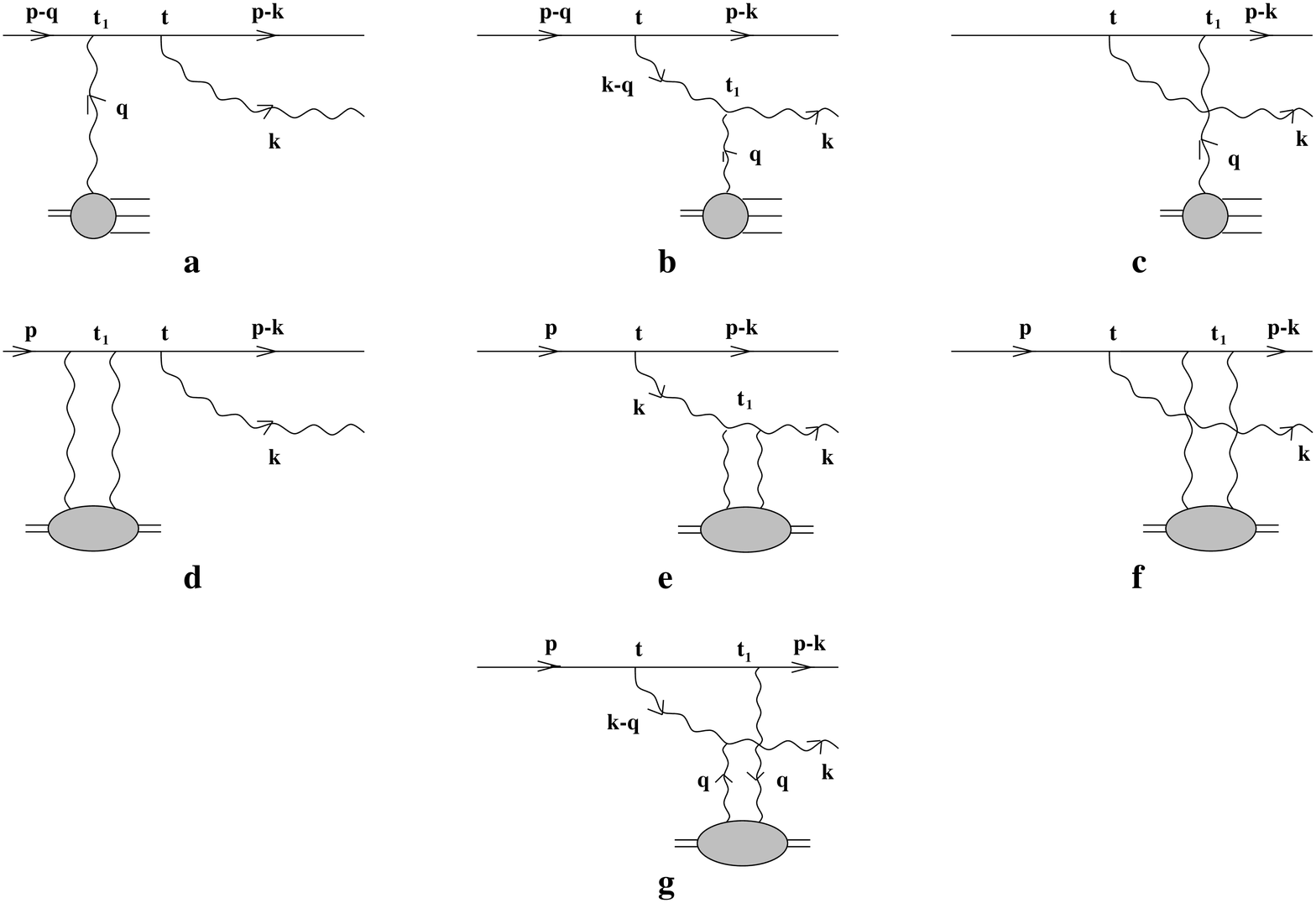,height=13.5cm,width=13cm}
\vskip .7cm
\caption{\label{fig:pict2} The diagrams representing the Born terms
for the amplitude. }
\end{figure}
Graphs a-c
correspond to inelastic reactions with the medium while graphs d-g correspond
to forward scattering in the medium. For terms a-c there are corresponding
inelastic reactions in the complex conjugate amplitude. In the approximation
that 
the forward elastic amplitude for quark scat\-te\-ring off particles in the
medium is purely imaginary, the elastic and inelastic terms are proportional to
$V(Q^2)$. The color factors and the expression of each graph contribution are
derived in \cite{Baier4}. 

The quark-gluon amplitude $f(\vec{U}, \vec{V} , t)$ obeys an integral evolution
equation derived in \cite{Baier1,Baier2,Baier4}.
 In impact parameter space and in the
small-$x$ limit, this equation takes the simple form
\begin{equation}
\label{17e}
{\partial \over \partial \tau} \widetilde{f}(\vec{B}, \tau ) = i
\widetilde{\kappa} \  \nabla_B^2 \ \widetilde{f}(\vec{B}, \tau ) - 2 (1 -
\widetilde{V} (B)) \widetilde{f}(\vec{B}, \tau ) \end{equation}

\noindent with $\widetilde{\kappa} = {2 C_F \over N_C} \left ( {\lambda \mu^2
\over 2 \omega} \right )$ and $\widetilde{f} 
(\vec{B}, 0) = \widetilde{f}_{Born}
(\vec{B})$. This equation is a Schr\"odinger-type evolution equation for the
propagation of the quark-gluon system in a QCD medium.
 Comparing (\ref{17e}) to  (\ref{11e}) is instructive. 
The term proportional to $\widetilde{\kappa}$
in (\ref{17e}) is clearly of quantum origin
and is  associated to the phase of the
amplitude whereas (\ref{11e}) is a classical diffusion equation. The
contributions entering the expression for the spectrum (\ref{14e}) are depicted
in Figure \ref{fig:pict7}.

\begin{figure}[ht]
\centering
\epsfig{file=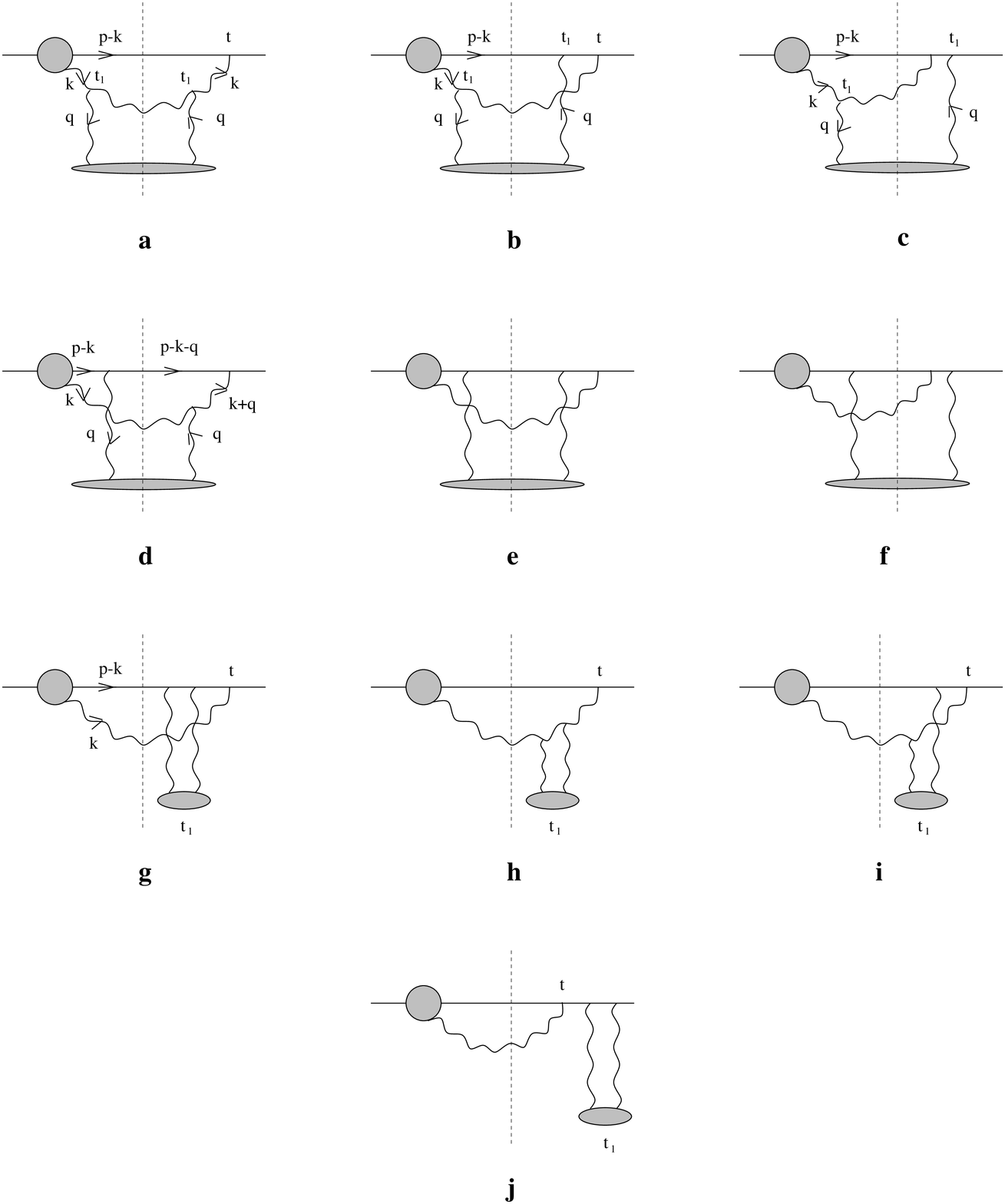,height=17cm,width=13cm}
\vskip .7cm
\caption{\label{fig:pict7}
Graphs describing the gluon emission.}
\end{figure}

So long as $\widetilde{v}(B^2) \equiv 4(1 - \widetilde{V}(B)/B^2$ can be
treated as a constant, solving (\ref{17e}) proceeds in analogy with
that of the 2-dimensional harmonic oscillator with imaginary frequency. We
expect that the behavior of $\widetilde{v}(B^2)$ is close in general to the
Coulomb potential case i.e. $\approx \, \ell n (1/B^2)$ at small $B^2$. 
The solution of
(\ref{17e}) to logarithmic accuracy
 is worked out in \cite{Baier2,Baier4}.

\noindent In the case where the (quark) jet is produced in matter, one finds
for the gluon spectrum
\beq
{\omega dI \over d \omega dz} = {2\alpha_s C_F \over \pi L}
[1 - x + {{x^2} \over 2} ] \ \ln \left |
\cos (\omega_0 \tau_0) \right | \> ,
\label{DiffSpectrum} 
\eeq 
from which the 
 energy loss per unit length is derived,
\begin{equation}
\label{18e}
- {dE \over dz} = \int_0^{\infty} {\omega \ dI \over d\omega \ dz},
\end{equation}
i.e.
\begin{equation}
\label{20e}
- {dE \over dz} = {\alpha_s \ N_c \over 4} \ {\mu^2L \over \lambda}
\widetilde{v}(\tau_0^{-1}) .
 \end{equation}

\noi
Notice the remarkable
 relation (cf. Eq. (\ref{eq:R16}))
\begin{equation}
\label{22e}
- {dE \over dz} = {\alpha_s \ N_c \over 4} p_{\bot W}^2
\end{equation}
\noindent between energy loss and jet $p_\perp$-broadening \cite{Baier3}.

\section{PATH INTEGRAL APPROACH}

This section is devoted to presenting in some detail a different approach
to the LPM effect in QCD, based on the so-called light-cone
 path integral technique for 
multiple scattering developed in \cite{Zakharov8}.
We refer the reader for derivations and further details to
\cite{Zakharova} - \cite{Zakharov3}.
Let us give here the essential features of the method by discussing
in scalar QED the formalism for an induced transition $a \rightarrow b c$. 
The interaction Lagrangian is $L_{int} = \lambda \left[ \hat\psi_b^{\dag}
 \hat\psi_c^{\dag}  \hat \psi_a + h.c. \right]$. 
It is assumed that the decay $a \rightarrow bc$
does not take place in the vacuum. We shall indicate later the proper 
treatment for realistic QED and QCD.

\def\rhoo{\vec{\rho}} 
\def\rr{\vec{r}}
\def\qq{\vec{q}}
\def\RR{\vec{R}}
\def\vtau{\vec{\tau}}
\def\vv{\vec{v}}

\subsection{Derivation of the basic formulas }

The $S$-matrix element for the $a\rightarrow bc$ transition in an external 
potential reads 
\begin{equation}  \label{eq:Z1}
\langle bc|\hat S | a \rangle = i \int dt d {\rr}~ \lambda(z)~  
\psi^*_b (t ,{\rr}) \psi^*_c (t,{\rr}) \psi_a (t ,{\rr}) ,
\end{equation}
where $\psi_i$ are the wave-functions, 
and $\lambda(z)$ adiabatically vanishes at $|z|\rightarrow \infty$ .
Let us consider the case of a static 
external potential. Then we can write $\psi_i$ as 
\begin{equation}  \label{eq:Z2}
\psi_i (t ,\rr) = \frac{1}{\sqrt{2E_i}} \exp [ -i E_i (t - z) ] \phi_i 
(z,  \rhoo),
\end{equation}
where $\rhoo$
 is the transverse coordinate, and the function $\phi_i$ describes
the evolution of the $\psi_i$ on the light-cone $t - z =\, {\rm const}$.
At high energies $E_i \gg m_i$, after substituting (\ref{eq:Z2})
into the Klein-Gordon equation, one can obtain for $\phi_i$ the 
two-dimensional Schr\"odinger equation
\begin{equation}  \label{eq:Z3}
i \frac{\partial\phi_i}{\partial z} = H_i (z) \phi_i , 
\end{equation}
\begin{equation}  \label{eq:Z4}
H_i (z) = - \frac{1}{2\mu_i} \left( \frac{\partial}{\partial\rhoo} \right)^2 + 
e_i U (\rhoo, z) + \frac{m^2_i}{2 \mu_i} , 
\end{equation}
where $\mu_i = E_i$, $ e_i$ is the electric charge, and $U$ is the potential
of the target. Consequently, the values of the $\phi_i$ at the 
$\rhoo$-planes $z = z_2$ and $z = z_1$ are related by
\begin{equation}  \label{eq:Z5}
\phi_i (z_2 , \rhoo_2 ) = \int d\rhoo_1 K_i (\rhoo_2 , z_2 | \rhoo_1 , z_1) 
\phi_i (z_1, \rhoo_1) , 
\end{equation}
where $K_i (\rhoo_2 , z_2 | \rhoo_1 , z_1 )$
is the Green function of the Hamiltonian (\ref{eq:Z4}).
 Let us introduce the two $\rhoo$-planes 
located at large distances in front of $(z = z_i)$ and behind $(z = z_f)$
the target. Then, using the convolution relation (\ref{eq:Z5}) one can 
express the incoming and outgoing wave-functions in terms of their asymptotic 
plane-waves at $z_i$ and $z_f$, respectively. As we shall see below, this 
representation turns out to be very convenient for the evaluation of the 
LPM effect. It is one of the key points of the light-cone path integral 
approach.

The differential cross section can 
be written as 
\begin{equation}  \label{eq:Z6}
\frac{d^5 \sigma}{dx d\qq_b d\qq_c} = 
\frac{2}{(2\pi)^4} {\rm Re} \int d\rhoo_1
d\rhoo_2 \, \int\limits_{z_1 < z_2}\, dz_1 dz_2 gF (z_1, \rhoo_1) F^* (z_2 , 
\rhoo_2),
\end{equation} 
where $F (z , \rhoo ) = \phi^*_b (z , \rhoo ) \phi^*_c (z , \rhoo ) \phi_a 
(z , \rhoo) $, $ \qq_{b,c}$ are the transverse momenta, 
$x = E_b / E_a$, and $g = \lambda(z_1)\lambda(z_2) / [16\pi x (1 - x) E^2_a]$ 
is the vertex factor.
Expressing  $\phi_i$ in terms of the asymptotic plane-waves,
 (\ref{eq:Z6}) may be represented diagrammatically by the graph of
Figure \ref{fig:bild}a. We
depict $K_i (K^*_i)$ by $\rightarrow (\leftarrow)$. The dotted lines show
the transverse density matrix for the initial particle $a$ at 
$z = z_i$, and the complex conjugate transverse density matrices for the 
final particles $b, c$ at $z = z_f$.
For the spectra integrated over transverse momenta, the 
relation
\begin{equation}  \label{eq:Z7}
\int d\rhoo_2 K (\rhoo_2 , z_2 | \rhoo_1 , z_1) K^* (\rhoo_2 , z_2 |
 {\rhoo_1}^{\,\prime} , z_1 ) = \delta (\rhoo_1 - {\rhoo_1}^{\,\prime})
\end{equation} 
allows one to transform the graph of Figure \ref{fig:bild}a
 into the ones of Figure \ref{fig:bild}b and 
Figure \ref{fig:bild}c for the $\qq_c$- and $\qq_{b,c}$-integrated spectra,
respectively.

\begin{figure}[ht]
\vskip 1cm
\centering
\epsfig{bbllx=36,bblly=283,bburx=526,bbury=369,
file=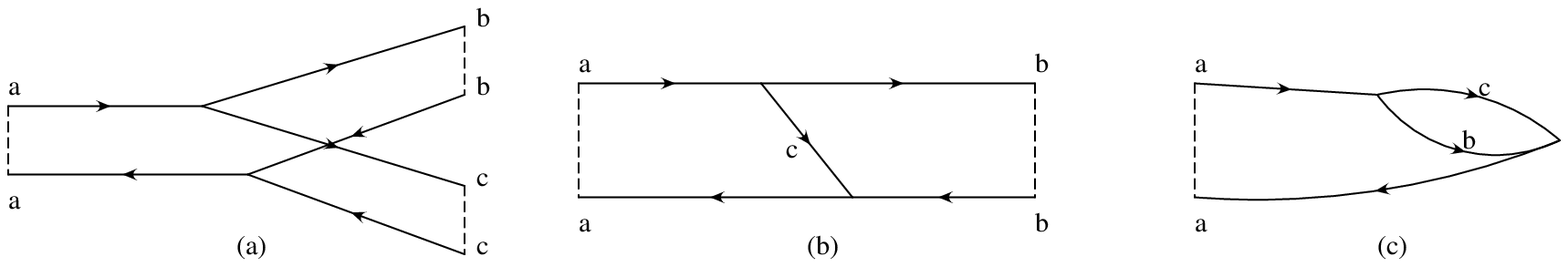, width=110mm,height=35mm}
\caption{\label{fig:bild}
Diagrammatic representation of the $a \rightarrow bc$ transition  
in terms of the two-dimensional Green functions.}
\vskip 1.5cm
\end{figure}

Let us discuss now the $a \rightarrow bc$ transition for a random potential of
an amorphous target using the representations of Figure \ref{fig:bild}.
 In this case one 
should perform averaging of the transition cross section over the states of 
the target. We cannot evaluate analytically the diagrams of 
Figure \ref{fig:bild} 
for a given state of the target. The basic idea of the approach 
of \cite{Zakharova,Zakharovb,Zakharov3} 
is to represent all the propagators in the path integral form 
\begin{equation}  \label{eq:Z8}
K_i (\rhoo_2 , z_2 | \rhoo_1 z_1 ) = 
\int  D \rhoo \, \exp \left\{ i  \int 
dz \left[ \frac{\mu_i (d\rhoo/dz)^2}{2} - e_i U (\rhoo , z) \right] - 
\frac{im^2_i (z_2 - z_1)}{2\mu_i} \right\} , 
\end{equation}
and to perform averaging over the target 
states before the integration over the trajectories. It is then remarkable 
 that  for the diagrams of Figure \ref{fig:bild}b,c 
(and for Figure \ref{fig:bild}a if $b$ or $c$ has zero
charge, say, for the $e \rightarrow \gamma e^\prime$ transition) a
considerable part of work on the path integration can be done analytically. 

Below we consider the $\qq_c$-integrated spectrum. Let $z_1$ and $z_2$ be 
the longitudinal coordinates of the left and right vertices of the 
diagram of Figure \ref{fig:bild}b.
 Taking advantage of the convolution relations (we omit 
the transverse variables)
\noindent
 $K_b (z_f | z_1) = K_b (z_f | z_2) \otimes K_b 
(z_2 | z_1)$ , $ K_a^* (z_2 | z_i) = K^*_a (z_2 | z_1 ) \otimes K^*_a 
(z_1 | z_i)$, 
\noindent
we can divide the diagram of Figure \ref{fig:bild}b 
into the initial and 
final state interaction two-body parts (we denote them by $S_c$ and $S_b$)
and the three-body part (we denote it by $M$) located between them. 
The factors $S_i$ and $M$ are given in terms of the Green functions
(\ref{eq:Z8}).

Let us first consider the factor $S_i$: 
after averaging over the states of the target, the phase factor
\begin{equation}  \label{eq:ZZ} 
\exp \{ i e_i \, \int \, dz [ U (\rhoo ,z) 
-U (\rhoo^{\, \prime} , z)]\}
\end{equation}
 can be viewed as the Glauber factor for the $i \bar i$
pair. Neglecting the correlations in the positions of the medium 
constituents one can obtain for the averaged value of this phase factor 
(we denote it by $\Phi_i (\{ \rhoo \} , \{ \rhoo^{\, \prime} \} )$)
\begin{equation}   \label{eq:Z12}
\Phi_i ( \{ \rhoo \} , \{ \rhoo^{\, \prime} \} )
 = \exp \left[ - \frac{1}{2} \int \,
dz\sigma_{i \bar i} (|\rhoo (z) - \rhoo^{\, \prime} (z) |) n (z) \right] ,
\end{equation}
where $\{ \rhoo \}$ and $\{ \rhoo^{\, \prime} \}$ are the trajectories for 
$K_i$ and $K^*_i$, respectively, $\sigma_{i \bar i}$ is the dipole 
cross section of the interaction with the medium constituent of the $i \bar i$
pair, and $n(z)$ is the number density of the target.
 Then the $S_i$ is 
given by the path integral $\int D\rhoo D\rhoo^{\, \prime} 
\exp [ i \hat{S}_i (\{ \rhoo \}, \{ \rhoo^{\,\prime} \} )]$ with the action
\begin{equation}  \label{eq:Z13}
\hat{S}_i ( \{ \rhoo \} , \{ \rhoo^{\, \prime} \}) = \frac{1}{2}\,\int\, dz \{
\mu_i [ (d\rhoo/dz)^2  - (d\rhoo^{\, \prime}/dz)^2 ]
 + i\sigma_{i \bar i} ( |\rhoo (z) - \rhoo^{\, \prime} (z) |) n (z) \} .
\end{equation}
It is important that the interaction term in (\ref{eq:Z13}) 
depends only on the relative distance between trajectories. This fact allows
one to carry out analytically the path integration and obtain a simple 
formula \cite{Zakharov8}
\begin{equation}  \label{eq:Z14}
S_i (\rhoo_2 , \rhoo_2^{\, \prime} , z_2 | \rhoo_1 , \rhoo_1^{\, \prime}, z_1 )
 = K^0_i (\rhoo_2 , z_2 | \rhoo_1 , z_1 )
 K^{0*}_i (\rhoo_2^{\, \prime}, z_2 | \rhoo_1^{\, \prime}, z_1 )
 \Phi_i (\{ \rhoo_s \} , \{ \rhoo_s^{\, \prime} \} ) , 
\end{equation} 
where
\begin{equation}  \label{eq:ZZ1}
K^0_i (\rhoo_2 , z_2 | \rhoo_1 , z_1 ) = \frac{\mu_i}{2\pi i (z_2 - z_1)} \,
\exp \, \left[ \frac{i \mu_i (\rhoo_2 - \rhoo_1)^2}{2(z_2 - z_1)}
 - \frac{im^2_i (z_2 - z_1)}{2\mu_i} \right] 
\end{equation}
is the Green function for $U = 0$, $\{ \rhoo_s \}$ and 
$\{ \rhoo_s^{\, \prime} \}$
denote the straight line trajectories between 
$\rhoo_{1,2}$ and $\rhoo_{1,2}^{\,\prime}$,
respectively. Expression (\ref{eq:Z14}) can be obtained by dividing the 
$z$-interval into steps of small width, and taking the multiple 
integral step by step \cite{Zakharov8}. 

The factor $M$  can be treated similarly. 
The corresponding Glauber factor contains the three-body cross 
section $\sigma_{\bar a b c}$ depending on the relative transverse 
separations $\vtau_{bc} = \rhoo_b - \rhoo_c$ and
 $\vtau_{\bar a c} = \rhoo_{\bar a} 
- \rhoo_c$ (here and below we view the particle $a$ for a complex conjugate 
propagator as antiparticle $\bar a$).
The path integrals may be performed analytically
leading to
\begin{equation}  \label{eq:Z15}
M(\rhoo_2, \rhoo_2^{\, \prime} , z_2 | \rhoo_1 , \rhoo_1^{\, \prime} , z_1 ) 
= K^0_a (
{\RR}_2 , z_2 | {\RR}_1 , z_1) 
K^{0*}_a (\rhoo_2^{\, \prime} , z_2 | \rhoo_1^{\, \prime}, z_1 ) 
K_{bc} (\rhoo_2 - \rhoo_2^{\, \prime} , z_2 | 0 , z_1 ),
\end{equation} 
where ${\RR}_1 = \rhoo_1$ , ${\RR}_2 = x \rhoo_2 + (1 - x) \rhoo_2^{\, \prime}$
are the initial and final coordinates of the centre-of-mass of the $bc$ 
pair, respectively. The Green function $K_{bc}$ is given by a path integral
on $\vtau_{bc}$ and describes the evolution of the $bc{\bar a}$ system.

Having (\ref{eq:Z14}) and (\ref{eq:Z15}), one can 
represent the spectrum for the $a\rightarrow bc$ transition for the diagram of 
Figure \ref{fig:bild}b for $\qq_a = 0$  in the form \cite{Zakharov3}
\begin{eqnarray}  \label{eq:Z16}
\frac{d^3 I}{dx d\qq_b} &=& \frac{2}{(2\pi)^2} {\rm Re} \int d \vtau_b \, 
\exp (-i \qq_b \cdot \vtau_b) \int\limits^{z_f}_{z_i} dz_1
\nonumber \\ &\cdot&  \int\limits^{z_f}_{z_1}  
dz_2 \,g \, \Phi_b (\vtau_b , z_2) K_{bc}(\vtau_b, z_2 | 0 , z_1) \Phi_a 
(\vtau_a , z_1),
\end{eqnarray}
where
\begin{eqnarray}  \label{eq:Z17}
\Phi_a ( \vtau_a , z_1 ) & =& \exp \left[ - \frac{\sigma_{a \bar a} 
(\vtau_a)}{2}\,
\int\limits^{z_1}_{z_i} \, dz  n (z) \right], \nonumber \\ 
\Phi_b (\vtau_b , z_2)& =& \exp \left[ - 
\frac{\sigma_{b \bar b} (\vtau_b)}{2} \, \int\limits^{z_f}_{z_2}
 dz n(z)\right] ,
\end{eqnarray}
are the values of the absorption factors for the parallel trajectories, 
and $\vtau_a = x \vtau_b$. The potential for the Green function $K_{bc}$
entering (\ref{eq:Z15}) and   (\ref{eq:Z16}) 
should be evaluated for parallel trajectories of the 
centre-of-mass of the $bc$ pair and $\bar a$. The resulting Hamiltonian for 
$K_{bc}$ is given by
\begin{equation}  \label{eq:Z18}
H_{bc} = - \frac{1}{2\mu_{bc}} \left( \frac{\partial}{\partial \vtau_{bc}}
\right)^2 - \frac{in(z)\sigma_{\bar a bc} (\vtau_{bc} , \vtau_{\bar a c})}{2}
+ \frac{1}{L_f} , 
\end{equation}
where $\mu_{bc} = E_a x(1 - x)$ is the reduced Schr\"odinger mass, 
$\vtau_{\bar a c} = x \vtau_{bc} - \vtau_a$ ;
the formation length is $L_f = 2 E_a x (1 - x) / [
m^2_b (1 - x) + m^2_c x - m^2_a x (1 - x) ]$. 

For numerical calculations it is convenient to represent (\ref{eq:Z16}) 
in another form in which
(for a target occupying the region $0<z<L$) 
the $z$-integration is dominated by the region
$|z|\leq \mbox{max}(L_f,L)$.
Let us rewrite the integrand of the $z_2$-integral in (\ref{eq:Z16}) as
\begin{eqnarray} \label{eq:bgz1}
& g&\{ \Phi_b(\vtau_b , z_2) [ K_{bc} (\vtau_b , z_2 | 0 , z_1) -
K^0_{bc} (\vtau_b , z_2 | 0 , z_1)] \Phi_a (\vtau_a , z_1 ) 
\nonumber \\
& +& [ \Phi_b (\vtau_b , z_2 ) - 1 ]
K^0_{bc} (\vtau_b , z_2 | 0 , z_1) 
[\Phi_a (\vtau_a , z_1 ) - 1 ]  
 + K^0_{bc} (\vtau_b , z_2 | 0 , z_1) 
[\Phi_a (\vtau_a , z_1 ) - 1 ] \nonumber \\
& +& [ \Phi_b (\vtau_b , z_2 ) - 1 ] K^0_{bc} (\vtau_b , z_2 | 0 , z_1)
 +  K^0_{bc} (\vtau_b , z_2 | 0 , z_1)\} .
\end{eqnarray}
It is evident that in (\ref{eq:Z16})
the contribution associated with the first two terms of (\ref{eq:bgz1}) 
will be dominated by the region $|z_{1,2}|\leq \mbox{max}(L_f,L)$, 
and can be evaluated 
neglecting the $z$-dependence of $\lambda(z)$.
However, for the last three terms in (\ref{eq:bgz1}) it is not the case.
Taking $z_{i}=-\infty$, $z_f=\infty$, and using $\lambda(z)$ which 
exponentially 
vanishes at $|z|\rightarrow \infty$ one can show that the term 
$\propto K^0_{bc}$ does not contribute to the spectrum. It is not surprising 
since this term corresponds to the transition in vacuum. The contribution
of the other two terms can be written in terms of the $bc$ Fock component 
of the light-cone wave-function of the particle $a$, $\Psi^{bc}_a$.
\footnote{These terms, which vanish after integrating 
over $\qq_b$,  have been missed in \cite{Zakharov3}.} 
The final expression for the spectrum is given by
\begin{eqnarray}  \label{eq:Z20}
\frac{d^3 I}{dxd\qq_b}& =&  \frac{2}{(2\pi)^2} {\rm Re}  \int  d\vtau_b \, 
\exp (- i\qq_b \cdot \vtau_b) \int\limits^{z_f}_{z_i}  dz_1  \int\limits^{z_f}_
{z_1} dz_2  
\nonumber \\
& \cdot &  g \, \{ \Phi_b
(\vtau_b , z_2) [ K_{bc} (\vtau_b , z_2 | 0 , z_1) - K^0_{bc} (\vtau_b , z_2 | 
0 , z_1)] \Phi_a (\vtau_a , z_1 ) \nonumber  \\
& +& [ \Phi_b (\vtau_b , z_2 ) - 1 ] K^0_{bc} (\vtau_b , z_2 | 0 , z_1) 
[\Phi_a (\vtau_a , z_1 ) - 1 ]\}\nonumber \\
& -& \frac{1}{(2\pi)^{2}}\int d\vtau d\vtau^{'}
\exp (- i\qq_b \cdot \vtau)
\Psi^{bc*}_a (x , \vtau^{\,'}-\vtau) \Psi^{bc}_a (x , \vtau^{\,'})\nonumber \\
& \cdot &\left[ \Phi_b (\vtau , z_i ) + \Phi_a (x\vtau , z_f ) - 2 \right] ,
\end{eqnarray} 
where one can take $z_i = - \infty , z_f = \infty$.

Integrating over $\qq_b$ one obtains from (\ref{eq:Z20}) the $x$-spectrum
\begin{equation}  \label{eq:Z21}
\frac{dI}{dx} = 2 {\rm Re} \int\limits^{z_f}_{z_i} dz_1 \int\limits^{z_f}_
{z_1}
dz_2 g \left. \left[K_{bc} (\rhoo_2 , z_2 | \rhoo_1 , z_1) - K^0_{bc} 
(\rhoo_2 , z_2 | \rhoo_1 , z_1 )\right]\right|_{\rhoo_1 = \rhoo_2 = 0} , 
\end{equation}
which was derived in \cite{Zakharova}
using the unitarity connection between the probability of the 
$a \rightarrow bc$ transition and the radiative correction to the 
$a \rightarrow a$ transition.

The spectrum (\ref{eq:Z21}) can be represented in another form
demonstrating a close connection between the LPM suppression and Glauber 
absorption. Treating the second term
of the Hamiltonian (\ref{eq:Z18}) as a perturbation, one obtains
\begin{eqnarray} \label{eq:Z22}
K_{bc} ( 0, z_2 | 0, z_1) & = & K^0_{bc} (0,z_2 | 0, z_1)
 + \int\limits^{z_2}_
{z_1} d \xi  \int d\rhoo \, K^0_{bc} (0, z_2 | \rhoo, \xi) v ( \xi, \rhoo)
\nonumber \\
\cdot \,  K^0_{bc} ( \rhoo , \xi | 0 , z_1) &+&
 \int\limits^{z_2}_{z_1}  d \xi_1 
\int\limits^{z_2}_{\xi_1} \, d\xi_2 \int  d\rhoo_1 d\rhoo_2
         K^0_{bc} (0, z_2 | \rhoo_2 , \xi_2) v ( \xi_2, \rhoo_2)
\nonumber \\
&\cdot&  K_{bc} (\rhoo_2, \xi_2 | \rhoo_1 , \xi_1) v ( \xi_1 , \rhoo_1)
K^0_{bc} (\rhoo_1 , \xi_1 | 0 , z_1) , 
\end{eqnarray}
where $v (z , \rhoo) = - n (z) \sigma_{\bar a bc} (\rhoo , x \rhoo)/2$. 
Taking advantage of (\ref{eq:Z22}) one can represent (\ref{eq:Z21}) in 
the form
\begin{equation}  \label{eq:Z23}
\frac{dI}{dx} = \frac{dI^{BH}}{dx} + \frac{dI^{abs}}{dx} , 
\end{equation} 
\begin{equation}  \label{eq:Z24}
\frac{dI^{BH}}{dx} = T \, \int \, d\rhoo \, | \Psi^{bc}_a (x , \rhoo) |^2 
\sigma_{\bar a bc} (\rhoo, x\rhoo ) , 
\end{equation} 
\begin{equation}  \label{eq:Z25} 
\frac{d I^{abs}}{dx} = - \frac{1}{2} {\rm Re} \int\limits^L_0  dz_1 n (z_1) 
\int\limits^L_{z_1} \, dz_2 n(z_2) \, \int \, d \rhoo \, 
\Psi^{bc*}_a (x, \rhoo ) 
\sigma_{\bar a bc} (\rhoo , x\rhoo ) \, \Phi (x, \rhoo , z_1 , z_2),
\end{equation}
where $T = \int^L_0 \, dzn (z)$ \cite{Zakharovb}, and
\begin{equation}  \label{eq:Zx5}
\Phi (x, \rhoo , z_1, 
z_2) = \int \, d\rhoo^{\, \prime} K_{bc} ( \rhoo , z_2 | \rhoo^{\, \prime}
 , z_1) \Psi^{bc}_a (x , \rhoo^{\, \prime} ) \sigma_{\bar a bc} 
(\rhoo^{\, \prime} , x \rhoo^{\, \prime} )
\end{equation}
is the solution of the Schr\"odinger equation with the boundary condition
\begin{equation}  \label{eq:ZZ5} 
\Phi (x, \rhoo , z_1 , z_1) = \Psi^{bc}_a (x , \rhoo ) \sigma_{\bar a bc}
(\rhoo , x \rhoo ). 
\end{equation}
The first term in (\ref{eq:Z23}) corresponds to the impulse 
approximation. It dominates the cross section in the low-density limit 
(the Bethe-Heitler regime). The second term describes absorption effects
responsible for the LPM suppression. 

For a sufficiently thin target the absorptive correction (\ref{eq:Z25}) 
can be evaluated neglecting the transverse motion in the $\bar a bc$ 
system inside the target (it corresponds to neglecting the kinetic term in 
(\ref{eq:Z18})). Then, the Green function takes a simple eikonal form
\begin{equation}  \label{eq:Z27} 
K_{bc} (\rhoo_2 , z_2 | \rhoo_1 , z_1) \approx \delta (\rhoo_2 - \rhoo_1 ) 
\exp \left[ - \frac{\sigma_{\bar a bc} (\rhoo_1 , x\rhoo_1)}{2} \,
\int^{z_2}_{z_1} \, dz n (z) \right] , 
\end{equation}
and (\ref{eq:Z23})-(\ref{eq:Z25}) give 
\begin{equation}  \label{eq:Z28}
\frac{dI^{fr}}{dx} = 2 \, \int \, d\rhoo \, | \Psi^{bc}_a (x , \rhoo ) |^2 
\Gamma^{eik}_{\bar a bc} ( \rhoo , x \rhoo ) , 
\end{equation} 
with $\Gamma^{eik}_{\bar a bc} (\vtau_{bc} , \vtau_{\bar a c}) = \{ 1 - \exp 
[ - T \sigma_{\bar a bc}(\vtau_{bc} , \vtau_{\bar a c}) / 2 ] \}$. 
This is the ``frozen-size'' approximation corresponding to the 
factorization regime discussed in section 3. 

The above analysis is performed for the particle $a$ incident on a target 
from outside. If it is produced in a hard reaction inside a medium one 
should replace in equations (\ref{eq:Z16}), (\ref{eq:Z20}) and (\ref{eq:Z21})
$z_i$ by the coordinate of the production point, and in (\ref{eq:Z20}) the 
factor $[\Phi_a - 1]$ should be replaced by $\Phi_a$. Note that due to 
the infinite time required for the formation of a light-cone wave-function 
$\Psi^{bc}_a$,  (\ref{eq:Z23}) does not hold in this case.

\subsection{Generalization to the realistic QED Lagrangian}

The generalization of the above analysis to the realistic QED and QCD 
Lagrangian is simple. Let us consider first the 
$e \rightarrow e^\prime \gamma$
transition in QED. The $\hat S$-matrix element can be obtained by replacing
in (\ref{eq:Z1}) 
$\lambda$ by $e\bar u_{e^\prime} \gamma^\nu \epsilon_\nu u_e$ 
where $\epsilon_\nu$ is the photon polarization vector, and $u_{e^\prime}, 
u_e$ are the electron spinors in which the transverse momenta should 
be regarded as operators acting on the corresponding wave-functions.
 Since the photon does not interact with the 
target, one has $\sigma_{\bar e \gamma e^\prime} (\vtau_{\gamma
e^\prime} , \vtau_{\bar e e^\prime} )= \sigma (| \vtau_{
\bar e e^\prime} |)$, where $\sigma$ is the dipole cross section
for the $e^+ e^-$ pair. In terms of the electron-atom differential cross 
section it reads
\begin{equation}  \label{eq:Z31}
\sigma (\rhoo) = \frac{2}{\pi} \int \, d\qq \, [ 1 - \exp (i\qq \cdot \rhoo)]
 \frac{d\sigma}{dq^2} .
\end{equation}
The dipole cross section vanishes as $\rhoo \rightarrow 0$, and one can write 
it as $\sigma (\rhoo) = C (\rho ) \rhoo^{\, 2}$, where $C (\rho)$ has a smooth 
logarithmic dependence at small $\rhoo$ \cite{Zakharov8,Nikolaev}.

In the Bethe-Heitler regime the radiation rate is dominated by $\tau_{\bar e
e^\prime} < 1 / m_e$; for the case of strong LPM suppression the 
typical values of $\tau_{\bar e e^\prime}$ are even smaller.
One can approximate the Hamiltonian (\ref{eq:Z18}) by the harmonic
oscillator Hamiltonian, and
obtain from (\ref{eq:Z21}) the radiation rate per unit length.
 In an infinite 
medium, in the regime of strong LPM suppression,
the radiation rate per unit length takes the form 
\begin{equation}  \label{eq:Z38}
\frac{dI}{dxdz} \approx \frac{\alpha [ 4 - 4x + 2 x^2]}{2\pi} \sqrt{
\frac{C(\rho_{eff} x)}{2x(1-x)E_e}} . 
\end{equation}
The value of $\rho_{eff}$ can be estimated as $\rho_{eff} \sim (2 L^\prime_
f / \mu_{\gamma e^\prime})^{1/2}$ where the formation length 
$L^\prime_f$ is the typical value of $|z_2 - z_1|$ in (\ref{eq:Z21}).
One can see that for strong suppression $\rho_{eff}$ becomes much smaller than
$1/m_e x$, the characteristic transverse size in the Bethe-Heitler regime. 
For this reason the spectrum for strong suppression (\ref{eq:Z38})
is insensitive to the electron mass \cite{BaierQed}.
 Note that the oscillator approximation
is equivalent to the Fokker-Planck approximation in 
momentum representation used in Migdal's analysis \cite{Migdal}.
This fact is not surprising since
within logarithmic accuracy  $\sigma (\rho) \propto \rhoo^{\, 2}$ leads to a 
Gaussian diffusion of the electron in transverse
 momentum space \cite{Zakharov8}.
This feature underlies the relationship given in section 3
relating the energy loss and $p_\bot$-broadening in QCD.

For an accurate numerical evaluation of the LPM effect it is convenient to 
use the form given by (\ref{eq:Z23})-(\ref{eq:Z25}).
 In \cite{Zakharovb,Zakharov7}
it is used for the analysis of the recent data on bremsstrahlung from high
energy electrons taken by the E-146 SLAC collaboration \cite{Anthony}.
Excellent agreement (at the level of 
the radiative corrections) with the data is found.

\subsection{Induced gluon emission in QCD}

Let us now discuss the induced gluon emission from a fast quark in QCD. 
At the level of the radiation cross section, involving the 
sum over states of the medium, one can formulate the theory 
similarly to the case of QED. The path integral representations for 
the diagrams of Figure \ref{fig:bild}
 can be written by introducing into the vacuum path 
integral formulas the Glauber factors for propagation of the color 
neutral partonic systems (consisting of the partons from the amplitude and 
complex conjugate one). The quark trajectory for the complex conjugate 
amplitude can be regarded as that of an antiquark with negative kinetic and 
mass terms. It follows from the relation $-T^*_q = T_{\bar q}$  (here 
$T_{q, \bar q}$ are the color generators for a quark and an antiquark). 
The $q \bar q$, $gg$, $q\bar q g$ configurations which can appear in the 
graphs like those of Figure \ref{fig:bild}b,c,
 have only one color singlet state, and 
the diffraction operator has only diagonal matrix elements involving the 
two-body cross sections $\sigma_{q \bar q} (\rhoo ) , \sigma_{gg}(\rhoo) =
\frac{9}{4} \sigma_{q \bar q} (\rhoo)$, and the three-body one $\sigma_
{gq\bar q} (\rhoo_{gq} , \rhoo_{\bar q q}) = \frac{9}{8} [ \sigma_{
q \bar q} ( | \rhoo_{gq}|) + \sigma_{q \bar q} (|\rhoo_{g\bar q}|)] - 
\frac{1}{8} \sigma_{q \bar q} (|\rhoo_{q \bar q}|)$
 \cite{Nikolaev}.
Thus, the spectra integrated over quark or/and gluon transverse momenta 
can be evaluated similarly to the above  case of the
$a \rightarrow bc$ transition in QED, with all the particles 
now being charged.

For the $x$-spectrum (\ref{eq:Z21}) $\rhoo_{\bar q q} = x \rhoo_{gq}$, 
and the three-body cross section takes the form $\sigma_{gq \bar q} ( \rhoo_
{gq} , x \rhoo_{gq}) = \frac{9}{8} [\sigma_{q \bar q} (\rho ) + 
\sigma_{q \bar q} ((1-x)\rho)] - \frac{1}{8} \sigma_{q \bar q} 
(x \rho)$, where 
$\rho = |\rhoo_{gq}|$. Similarly to QED, one can estimate the spectrum 
using the oscillator parametrization $\sigma_{gq\bar q} (\rhoo , x\rhoo) 
\approx C_3 (x) \rhoo^{\, 2}$,
 where $C_3 (x) = \frac{1}{8} \{ 9 [ 1 + (1 - x)^2]
-x^2\} C_2 (\rho_{eff}), \,  C_2 (\rho_{eff}) = \sigma_{q \bar q} 
(\rho_{eff}) / \rho^{\, 2}_{eff}$.
 Here $\rho_{eff}$ is the typical size of the 
$q \bar q g$ system dominating the radiation rate, which  in the 
limit of strong LPM suppression takes the form
\begin{equation}   \label{eq:Z42}
\frac{dI}{dxdz} \approx \frac{\alpha_s (4 - 4x + 2x^2)}{3\pi} 
\sqrt{ \frac{2nC_3(x)}{E_q x^3 (1 - x)}} . 
\end{equation}
Ignoring the contributions to the energy loss from the two narrow regions 
near $x \approx 0$ and $x \approx 1$, in which (\ref{eq:Z42}) is not 
valid, one finds the energy loss 
per unit length 
\begin{equation}  \label{eq:Z43}
\frac{d \Delta E_q}{dz} \approx 1.1\alpha_s \sqrt{ nC_3 (0) E_q} ,
\end{equation}
where to logarithmic accuracy $\rho_{eff} \sim [ \alpha^2_s n E_q x 
(1 - x )]^{-1/4}$ is taken. 
Note that as in QED, the elimination of the infrared 
divergence for strong suppression is a direct consequence of the medium 
modification of the gluon formation length which makes the typical 
transverse distances much smaller than $1 / m_{g, q}$.

The medium modification of the formation length plays an important role in 
the case of gluon emission by a quark produced \underline{inside} a medium. 
In this case the finite-size 
effects become important and suppress the radiation rate
 (cf. (\ref{DiffSpectrum})).
This finite-size suppression leads to the $L^2$ dependence of the quark
energy loss for a high energy quark (\ref{20e}). 
One obtains
\begin{equation}  \label{eq:Z46}
\Delta E_q \sim \alpha_s C_3 (0) n L^2 .
\end{equation}
This regime takes place as long as $L \le (E_q / n C_3 (0))^{1/2}$. 
Then it transforms into the $\Delta E_q \propto L$ behavior given by 
(\ref{eq:Z43}).
More detailed discussions and numerical estimates 
 are given in \cite{Zakharov5,Zakharov6}. 
The $g\rightarrow gg$ transition
can be evaluated in an analogous way. The $ggg$ system can be
in antisymmetric ($F$) and symmetric ($D$) color states.
However, the two-gluon Pomeron exchanges do not
generate the $F\leftrightarrow D$ transitions. This allows one
to express the emission probability through the Green
function for $F$ state. 

\subsection{Comparison with the BDMPS approach}

We conclude this section with a comment on the connection between the path 
integral approach with the approach discussed in section 3.

Let us consider the case of a parton entering the medium 
from outside. 
The equivalence of the two approaches may be established 
using (\ref{eq:Z23})-(\ref{eq:Z25}) together with (\ref{eq:Z28}),
in the zero mass case as assumed in BDMPS \cite{Baier2}.
 As it was mentioned, the "frozen-size" 
expression (\ref{eq:Z28}) corresponds to the factorization contribution,
neglected in the BDMPS approach, on the ground of its weak medium dependence.
Rewriting (\ref{eq:Z23}) as
\begin{equation}  \label{eq:Z48}
\frac{dI}{dx} = \frac{dI^{abs}}{dx} + \frac{dI^{fr}}{dx} - 
(\frac{dI^{fr}}{dx} -  \frac{dI^{BH}}{dx}), 
\end{equation} 
 and ignoring the second term, one can show that 
\begin{equation}  \label{eq:Z49}
\frac{dI}{dx} = \frac{dI^{abs}}{dx} {\Bigg|}^{\omega}_{\omega = \infty},
\end{equation}
together with identifying in (\ref{eq:Z25})
 the product $\Psi^{bc}_a \sigma_{\bar a bc}$ and  $\Phi$ 
 with the amplitudes $f_{{\rm Born}}$ and 
$f$, respectively, which are discussed in section 3.

For the case of a parton produced inside a medium, say at $z = 0$, in 
(\ref{eq:Z21}) $z_i = 0$, one should subtract from the right hand side 
of (\ref{eq:Z23}) the contribution corresponding to the configurations 
with $z_1 < 0$ and $z_2 > 0$ in (\ref{eq:Z21}). The 
additional term corresponds to the additional contribution in the 
BDMPS approach due to the hard scattering in the medium.
In this case the "frozen-size" expression is medium independent
and it may be obtained by taking the limit
of $dI^{abs}/dx$ when $\omega \rightarrow \infty$.
Further details concerning this comparison can be found in  \cite{Baier4}.

\section{RADIATIVE ENERGY LOSS IN AN EXPANDING QCD PLASMA}

In the previous sections we have discussed 
the  suppression of  gluon radiation 
due to multiple scatterings of energetic
partons propagating through dense matter with properties
constant in time.

Here we consider the case of a  parton, of high
energy  $E$, traversing an \underbar{expanding} hot QCD medium.
We concentrate on the
 induced  gluon radiation, the resulting  
 energy loss of a quark, and its relation to jet
broadening \cite{Baier5,Zakharov}.
 
 Let us imagine the medium to be a quark-gluon
plasma produced in a relativistic
central AA collision, which occurs at (proper) time $t = 0$.
We have in mind the realistic situation
where the quark is produced by a hard scattering \underbar{in} the (not yet
thermalized) medium, and
 at time $t_0$ it enters the homogeneous plasma at high 
temperature $T_0$, which expands longitudinally with respect to the 
collision axis.
Consider $t_0$ to be the thermalization time,
and for most of the results the limit $t_0 \rightarrow 0$ may be taken with
impunity. 
 The quark, for simplicity, is assumed to propagate in the 
transverse direction with vanishing longitudinal momentum,
i.e. at rapidity $y = 0$, such that its 
 energy is equal to its transverse momentum.
 On its way through the plasma 
the quark hits layers of matter which are cooled down due 
to the longitudinal expansion. It is  assumed that the plasma lives long 
enough so that the quark is able to propagate on a given distance 
$L$ within the quark-gluon  phase of matter.

As a consequence  of the medium expansion the  parton  propagation 
in the transverse direction, $z$, is affected by the position-dependent 
density of the plasma $\rho(z)$ and the parton cross 
section $d\sigma / d^2 \vec q_\bot (\vec q_\bot , z)$. 
Therefore the screening mass $\mu$ and the mean free path
$\lambda$ depend on $z$.
When the properties of the expanding plasma are described by the 
hydrodynamical model proposed by Bjorken \cite{Bjorken},
one has the scaling law
\begin{equation}   \label{eq:1.1}
T^3  t^\alpha = {\rm const} , 
\end{equation}
where the (proper) time $t$  at 
rapidity $y = 0$  coincides with the distance  $z$ on which
the quark has propagated through the plasma. The power $\alpha$,  
approximated in the following by a constant,
 may take values between $0$  and $ 1$ for an ideal fluid.

Correspondingly, the transport coefficient $ \hat q (t)$
defined as 
\begin{equation}\label{eq:2.9}
\hat q (t) 
 \simeq  \,\,
 \rho (t) \int d^2 \vec q_\bot \vec q_\bot^{~2} \frac{d\sigma}
{d^2 \vec q_\bot} = \frac{\mu^2(t)}{\lambda(t)} \tilde{v}
\end{equation}
 becomes  time-dependent and  satisfies
\begin{equation}   \label{eq:3.88}
\hat q (t) = \hat q (t_0) \left( \frac{t_0}{t} \right)^\alpha ,
\end{equation}
due to (\ref{eq:1.1}).

 As a result \cite{Baier5} the radiative energy loss $  \Delta E$
for the quark (produced \underbar{in} the medium)
traversing an expanding medium  is  

\begin{equation}     \label{eq:4.11}
- \Delta E = 
  \frac{2}{2 - \alpha } \frac{\alpha_s N_c}{4} \, 
\hat q (L) L^2 .
\end{equation}
In the high temperature phase of QCD matter 
\cite{Nieto} 
\begin{equation}  \label{eq:2.17}
1 - \alpha  = O ( \alpha^2_s (T) ) .
\end{equation}
\noi
The coefficient $\hat q (L) \, = \, \hat q (T (L))$ has to be 
evaluated at the 
temperature $T (L)$ the quark finally ``feels'' \underbar{after} 
having passed the distance $L$ through the medium, which during this 
propagation cools down to $T(L)$.
One may, however,  notice that the limit $\alpha = 1$
for an expanding ideal relativistic plasma can be taken. In this limit the
maximal loss is achieved. It is bigger by a factor 2  than the 
corresponding static case
at fixed temperature $T (L)$.
 
So far we have discussed the result for the case for $E > E_{cr}(L)$,
actually taking $E \to \infty$. In \cite{Zakharov} the approach 
of the quark's energy loss $\Delta E$ to this limit is studied numerically
as a function of the quark energy $E$. For instance, with
$L = 6$ fm, one finds  (almost) energy
independence on $E$, when $E > 100$ GeV $ \simeq E_{cr}$,
as given by (\ref{eq:4.11}).


 In summary one expects indeed that the energy loss in an expanding medium
 be larger than in the static case taken at the final temperature,
 since the parton passes through hotter layers during the early
 phase of the expansion. Perhaps the surprising feature is that
 there is no dependence of the enhancement factor on the initial
 temperature $T_0$.
 This result has to be associated to the coherence
 pattern of the medium induced radiation. Gluons contributing
 to the energy loss require finite time for their emission, and
 therefore effects of the early stages of the quark-gluon plasma
 expansion are reduced.

\section{INDUCED ENERGY LOSS OF A HARD QUARK JET IN A FINITE CONE}
 
Let us consider a typical calorimetric measurement of hard jets
 produced in heavy ion collisions \cite{Lokhtin1}.
 The consequence of a large
energy loss is the attenuation of the spectrum usually denoted as jet
quenching. It is necessary to study the angular distribution of radiated gluons
in order to give quantitative predictions for the energy
 lost by a jet traversing hot matter. Only the gluons
 which are radiated outside the cone defining the jet
contribute to the energy loss. \par

In \cite{Zakharov3,Baier6,Wiedemann} the calculation of
 the angular distribution is discussed for
 a hard jet produced in the medium.
Here we have in mind a hard
quark jet of high energy $E$ produced by a hard 
scattering in a dense QCD medium and propagating through it over a 
distance $L$. 
Following \cite{Baier6} we concentrate on the integrated loss 
{\it outside} an angular cone of opening angle $\theta_{{\rm cone}}$
(Figure \ref{fig:graph3}),
\begin{equation}\label{eq:3.1}
- \Delta E (\theta_{{\rm cone}}) = L \, \int^\infty_0 \, d\omega\,
\int^\pi_{\theta_{{\rm cone}}} \, \frac{\omega dI}{d\omega dz d\theta} 
d\theta .
\end{equation}

\begin{figure}[ht]
\centering
\epsfig{file=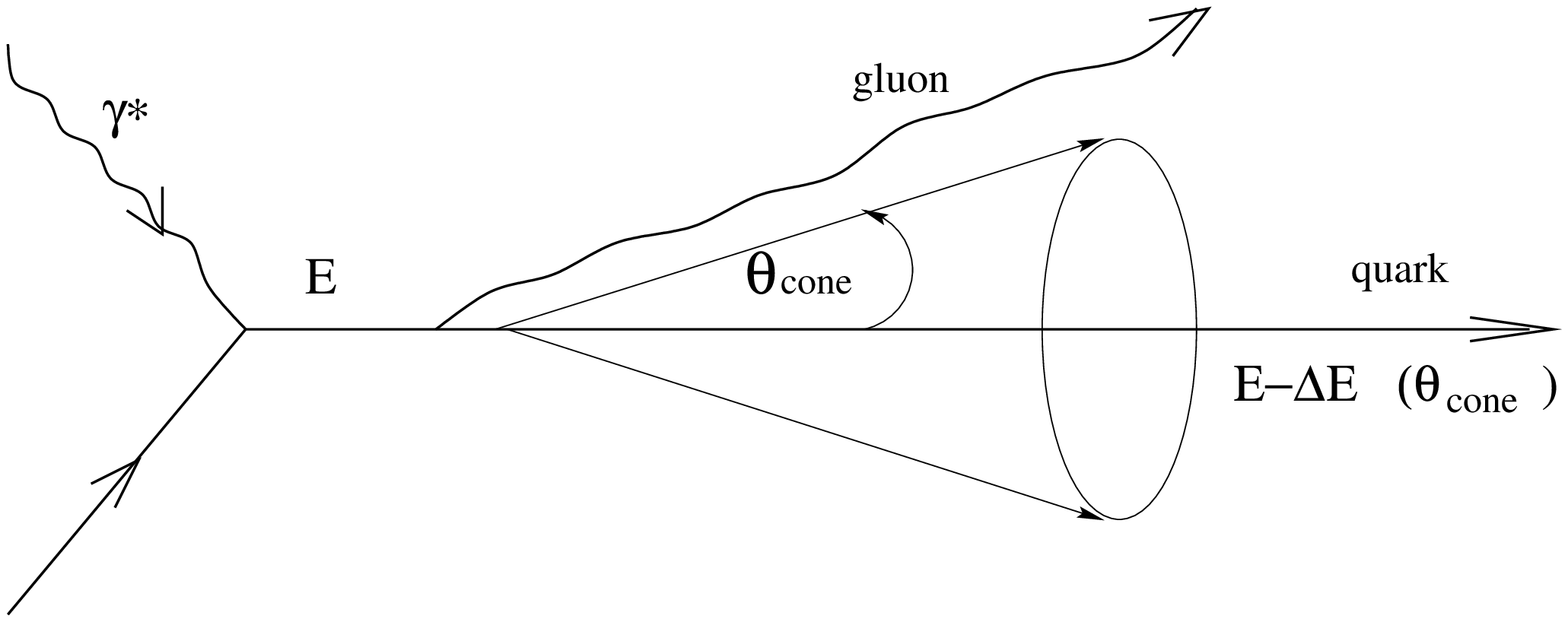,angle=-90,width=9cm}
\vskip  1.5cm
\caption{\label{fig:graph3} 
Example of a hard process producing a quark jet. The gluon is emitted outside
the cone with angle $\theta_{{\rm cone}}$ . }
\vskip  1.0cm
\end{figure}
In the following we consider the normalized loss by defining the ratio 
\begin{equation}\label{eq:3.3}
R (\theta_{{\rm cone}} ) = \frac{\Delta E (\theta_{{\rm cone}})}
{\Delta E}. 
\end{equation}
This ratio $R (\theta_{{\rm cone}})$
turns out to depend  on \underline{one} single dimensionless variable
\begin{equation}\label{eq:3.8}
R = R (c (L) \theta_{{\rm cone}} ), 
\end{equation}
where
\begin{equation}\label{eq:3.9}
c^2 (L) = \frac{N_c}{2C_F} \hat q \left( L / 2 \right)^3 .
\end{equation}

The ``scaling behaviour'' of $R$
means that the medium and size dependence is universally contained in the
function $c (L)$, which is a function of the transport coefficient $\hat q$ 
and of the length $L$, as defined by (\ref{eq:3.9}). 
In Figure \ref{fig:graph5},
 we show the variation of $R$ with $\theta_{cone}$.
\begin{figure}
\centering
\vskip 1cm
\epsfig{bbllx=45,bblly=205,bburx=510,bbury=610,
file=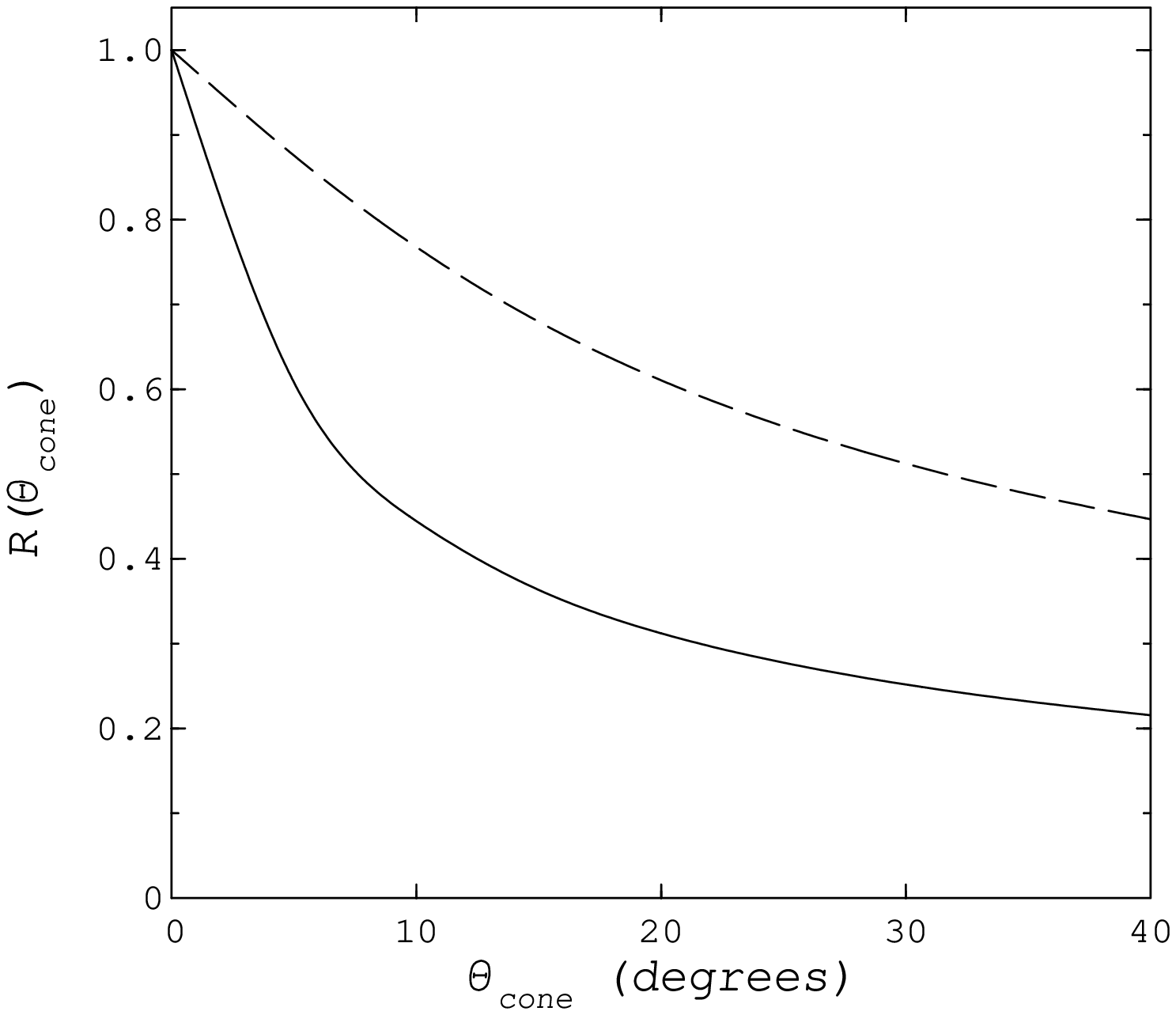,width=96mm,height=80mm}
\caption{\label{fig:graph5}Medium induced (normalized) energy loss 
distribution as a function of cone angle $\theta_{{\rm cone}}$ for hot 
($T=250$ MeV) (solid curve) and cold matter (dashed curve) at fixed 
length $L=10$ fm.}
\vskip 1.5cm
\end{figure}
The ratio $R(\theta_{{\rm cone}})$ is also universal in the sense
that it is the same for an energetic quark as well as for a gluon jet.
The fact that $\theta_{{\rm cone}}$ scales as $1/c(L)$ may be understood
from the following physical argument \cite{Dokshitzer}:
the radiative energy loss of a quark jet is dominated by gluons having 
$\omega \simeq \hat q L^2$. 
 The angle that the emitted gluon makes with the quark is $\theta \simeq 
{{k_{\perp}}}/{\omega}$, and  
$k_{\perp}^2 \simeq \hat q L$ so that
 the typical gluon angle will be $\theta^2 \simeq 1/{\hat q L^3}$.

So far we have discussed the {\it medium-induced} energy loss.
 Concerning the total energy loss of a jet of a given
cone size it is 
important to take into account the {\it medium independent} part,
which for a quark jet in a 
cone may be estimated \cite{Baier6}, 
\begin{equation}\label{eq:5.10}
 - \Delta E^{fact} ( \theta_{{\rm cone}}) 
\simeq \frac{4}{3} \frac{\alpha_s C_F}{\pi} 
E \, \ln \left( \frac{\theta_{{\rm max}}}{\theta_{{\rm cone}}}\right)
, 
\end{equation}
using a constant $\alpha_s$
 $(\theta_{{\rm max}}$ is 
taken ${\cal O} (\pi/2))$.

\section{PHENOMENOLOGICAL IMPLICATIONS}
  
The parameter controlling the magnitude of the energy loss is $\widehat{q}$.
Estimates can be provided for its value,
 allowing us to give orders of magnitude
for the radiative induced energy loss. The following numbers are
estimates for a quark jet produced in matter.

For \underline{hot matter}
taking $T =$ 250 MeV, ${\mu^2 \over \lambda} \sim 1$~GeV/fm$^2$ taken from
perturbative estimates at finite $T$,
 a typical value for $\widetilde{v} \approx
2.5$, we find $\widehat{q} \simeq 0.1$~GeV$^3$ \cite{Baier3}.
 With $\alpha_s = {1
\over 3}$, this leads for the total induced energy loss to
\begin{equation}
\label{23e}
- \Delta E \approx 60 \ {\rm GeV} \ \left ( {L \over 10 \ {\rm
fm}} \right )^2\quad . 
\end{equation}

In \cite{Baier3} it is  shown that for \underline{cold nuclear matter}
 it is possible to relate $\widehat{q}$ to the
gluon structure function $G$ evaluated at an average scale
 $\mu^2 {\lambda \over L}$, actually
\begin{equation}
\label{transp}
\widehat{q} \simeq {{2 \pi^2 \alpha_s} \over 3} \rho \, [ x G(x)].
\end{equation}
 Taking the nuclear density $\rho \sim 0.16$~fm$^{-3}$,
$\alpha_s = {1 \over 2}$, $xG \sim 1$ for $x < 0.1$, it is found that
\begin{equation}
\label{24e}
- \Delta E \approx 4 \ {\rm GeV} \ \left ( {L \over 10 \ {\rm fm}}
\right )^2 \quad .
\end{equation}

\noindent These values do
suggest that hot matter may be effective in stimulating 
significant radiative energy loss of high energy partons.
As discussed in section 5 the energy loss is larger in an expanding
hot medium than in the corresponding static one.

Next we turn to  the medium-induced $-\Delta E
 (\theta_{{\rm cone}})$ for energetic jets.
We may use the estimates above to give orders of magnitude for $c(L)$ in
the case of a \underline{hot/cold} medium~: 
$$c(L)_{hot} \simeq 40 \ (L/10 \ {\rm fm})^{3/2} \quad .$$

\noindent A much smaller value is found in the cold nuclear matter case~:
$$c(L)_{cold} \sim 10 \ (L/10 \ {\rm fm})^{3/2} \quad .$$

\noi
 As
expected from the fact that $R(\theta_{cone})$ depends universally
 on $c(L) \theta_{cone}$, 
Figure \ref{fig:graph5}
 shows that the jets are more collimated in the hot medium
 than in the cold one. 
The loss is, however, still appreciably large  even for cone sizes of order
 $\theta_{{\rm cone}} \simeq 30^\circ$.

\noi
Again,
keeping in mind that the  estimates are 
based on the leading logarithm approximation, we show
 in Figure \ref{fig:graph8} the variation of $\Delta E (\theta_{cone})$
 with $\theta_{cone}$ of
the medium-induced (for a hot medium with $T = 250$ MeV) and the medium 
independent energy losses.

\begin{figure}
\centering
\epsfig{bbllx=30,bblly=205,bburx=495,bbury=600,
file=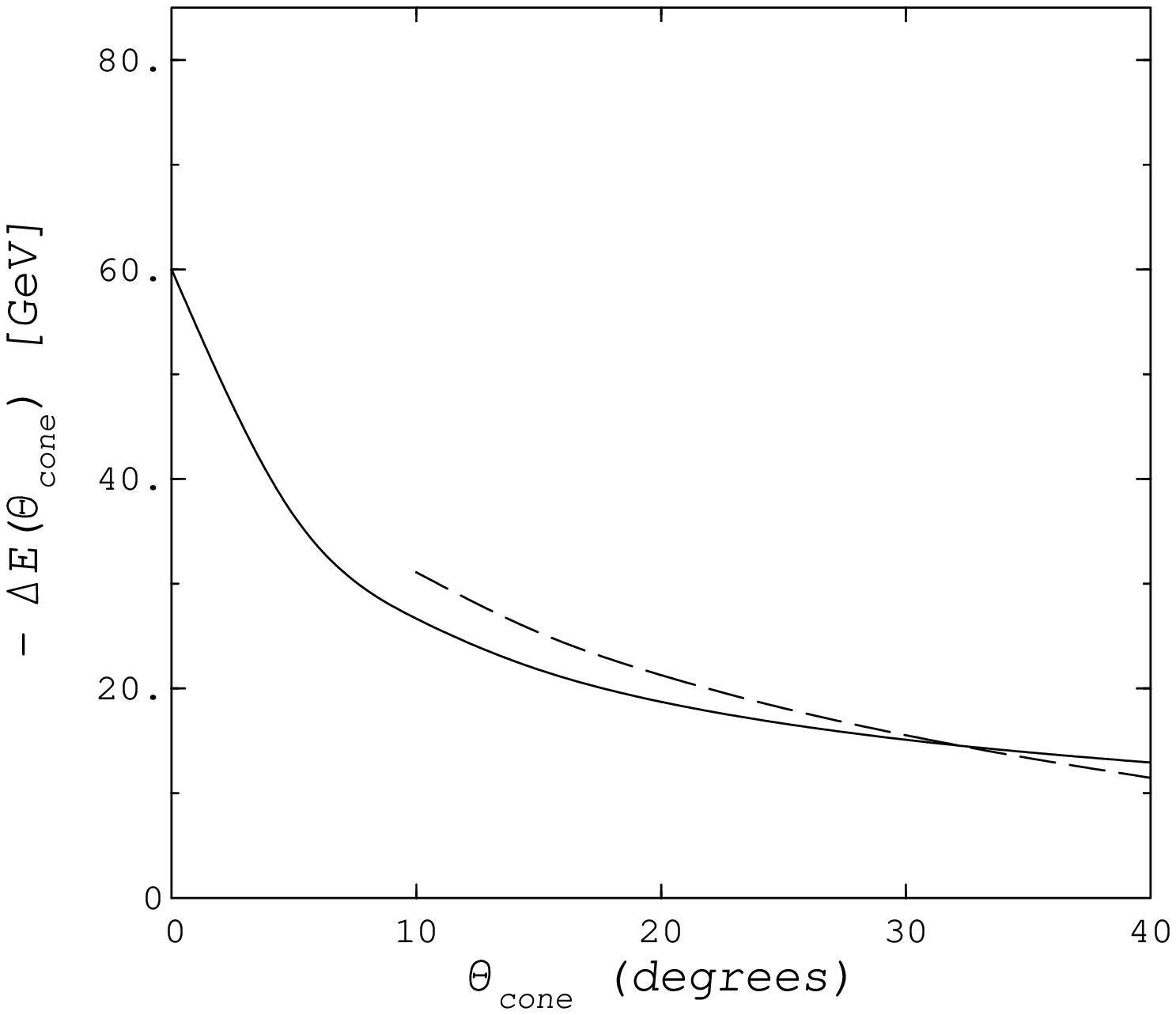,width=88mm,height=80mm}
\caption{\label{fig:graph8} Energy loss in a hot medium, $T=250 $ MeV,
as a function of $\theta_{{\rm cone}}$.
The dashed curve represents the medium independent piece
 for $E=250$ GeV. $L=10$ fm.  }
\end{figure}

Let us now give a few representative examples
of phenomena sensitive to parton energy loss in dense matter.
Available  experimental results are, it seems,
essentially instructive for future measurements at higher energies.

As a projectile traverses dense nuclear matter the width of the
transverse momentum distribution of partons may increase.
pA and AA scattering allow to study parton $p_{\bot}$-broadening 
of initial quarks in the Drell-Yan process of lepton pair
production, and of gluons in $J/\Psi$ production,
respectively (see e.g. \cite{Gavin,Hufner}). Recently,
the analysis of $J/\Psi$ data shows indeed $p_{\bot}$-broadening
 of the intrinsic gluon distribution, which when translated into the
ratio $\widehat{q}/\rho$ results into $\widehat{q}/\rho = 9.4 \pm 0.7$
\cite{Kharzeev}. Neglecting final state effects this should be compared 
with $\widehat{q}/\rho \sim 7.4 \, [x G(x)]$.

The above quoted processes  also contain information on the energy
loss in the initial state due to matter effects \cite{Brodsky,Nagle}.
In the Drell-Yan process the observed energy loss $ -\Delta E$
of the incident quarks is indeed compatible with the estimate given in
(\ref{24e}), including the $L^2$ dependence,
as measured and analysed in \cite{Vasiliev}.

Large $p_{\bot}$ particle and jet spectra and  production rates
in high-energy collisions are especially sensitive to a finite energy loss,
when the partons are propagating through long-lived high density media before
hadronization. Under extreme conditions the jets may even be "extincted"
\cite{Bjorkena}. However, hadron spectra from present experiments
of pp, pA and AA collisions,
mainly from CERN-SPS, do not show any strong evidence of suppression
\cite{Wangb,Wangc}. This observation
which is obscured by large theoretical uncertainties may indicate
that at present energies the life time of the dense
 partonic matter may be shorter than
the mean free path of the propagating partons.
Significant jet quenching should become
 clearly observable in AA collisions at RHIC
 and higher energies, even for transverse momenta as low
as $p_{\bot} \ge 3$ GeV:
the magnitude of the predicted jet quenching is
commented upon in \cite{Miklos,Mgyulassy}.
Suppression of hadronic $p_{\bot}$ distributions in the case of  a thin plasma,
therefore  due to only a small number ($\le 3$) of  scatterings, is analysed
 in \cite{Vitev}.

Jet quenching for very high energy jets 
 is also discussed in \cite{Lokhtin1,Lokhtin2,Lokhtin3}.
In particular the ratio of 
monojet versus  dijets observed in ultra-relativistic heavy ion collisions
is predicted.

A further interesting proposal to study the modification of jet fragmentation
due to energy loss is proposed in \cite{Wangd,Wange}. Noting that photons
are essentially not affected by hadronic media, the conjecture is
to measure the charged particle $p_{\bot}$ distribution in the opposite
transverse direction of a tagged photon, i.e. in $\gamma ~+$ jet 
events of high energy heavy ion collisions.
 With increased luminosity this may be even possible at RHIC energies.

Valuable and important information about dense hadronic matter produced
in collisions is provided by dileptons, either from
Drell-Yan processes or from final heavy meson decays \cite{McGaughey}.
In this context Shuryak \cite{Shuryak} pointed out the importance of the
energy loss of charm (bottom) quarks due
to their interactions in the medium. In the extreme case they may be even 
stopped in dense matter. Assuming e.g.
that the charmed mesons $D$ and $D^*$ take all the charm quark momentum in the
fragmentation process the final leptons from the semileptonic decay populate
the invariant dilepton mass spectrum at masses below $1 - 2$ GeV
 ($4 - 5 $ GeV from bottom decays):
as a result dilepton spectra in AA collisions for invariant masses
above $2$ GeV are {\underline{not}} dominated by correlated semileptonic charm
and bottom decays. This expected 
strong suppression due to energy loss is confirmed
in further detailed (Monte Carlo) studies in \cite{Lin1,Lin2,Gallmeister}.
 
\section{OUTLOOK}

In this review we have described the more recent results
related to  energy loss and $p_{\bot}$-broadening
of a high energy quark or gluon (jet) traversing QCD media.
Phenomenological implications for measurements in cold as well
in hot (QGP) matter have been discussed.  
The orders of magnitude found for the energy lost by an energetic jet
in hot deconfined matter indicate the interest of the corresponding
measurements as specific signals.

A couple of important open questions triggered by the coherent
character of the induced energy loss remain open. One is 
related  to the  formulation of a 
transport model (Monte Carlo) which 
correctly simulates the interference pattern
of gluon radiation  induced by multiple scattering 
 \cite{Gyulassy,Dokshitzer}.
It is indeed crucial when calculating rates for processes
 leading to thermal and chemical
equilibration of partons to include medium effects
\cite{Biro} - \cite{Eskola}.
In the same context let us mention the influence of the LPM effect on
the production of dileptons and real photons produced in a QGP,
or in a hadron gas \cite{Cleymans}. In any case
the partons are not very energetic, 
since their  energies are of the order of the plasma temperature.
This forbids to use the asymptotic treatment discussed in the above.

We already mentioned  that the above discussed  numerical estimates
give only orders of magnitude in particular
since  they are obtained in leading order in the QCD coupling.
The main aim of the present investigations  is therefore 
 to encourage experimentalists at RHIC, and later at LHC,
to carefully explore heavy quark production and especially
 jet phenomena in ultra-relativistic heavy ion
collisions \cite{Harris}.
High $p_{\bot}$ nuclear physics may become an exciting new frontier at
these colliders \cite{Mgyulassy}, because of the 
possible jet "extinction" or crucial modifications
of the spectra \cite{Bjorkena}.
 The best guiding example  comes from the long-term  study of hard jets
in hadron-hadron scattering, starting from the
first evidence at the CERN-ISR until the analysis of jet cross sections
up to transverse momenta of $p_{\bot} \simeq 500$ GeV
at CDF and D{\O}, which has been successfully 
carried out within an active  interplay
between experiments and  perturbative QCD \cite{Sterman}. 

Medium effects will continue to attract increasing attention, and we
hope that finally the described suppression mechanism will be
demonstrated by future experiments in a convincing manner.
More detailed treatments and improvements are certainly necessary
in view of completing this program. It constitutes an important
chapter of what has been recently designated \cite{Dokshitzer2}
as the "health report" of QCD.

\section{ACKNOWLEDGEMENTS}

R.~B. and D.~S. are grateful for the pleasant and fruitful collaboration
with Yu.~L.~Dokshitzer, A.~H.~Mueller and S.~Peign{\'e}
on the different aspects of these topics during the
last few years.
We thank M.~Dirks, I.~P.~Lokhtin, D.~Denegri,
K.~Redlich, H.~Satz, E.~Shuryak and A.~Smilga  
for valuable comments and discussions.
Partial support by DFG under contract Ka 1198/4-1
is acknowledged.

\end{document}